\documentclass[a4paper,fleqn,usenatbib]{mnras}

\usepackage{newtxtext,newtxmath}

\usepackage[T1]{fontenc}
\usepackage{ae,aecompl}
\usepackage{bm}

\usepackage{graphicx}

\usepackage{multirow}

\title[Limits on repeating FRBs]{Limits on the population of repeating fast radio bursts from the ASKAP/CRAFT lat50 survey}

\author[C. W. James]{
C.\ W.\ James,$^{1}$\thanks{E-mail: clancy.james@curtin.edu.au (CWJ)}
\\
$^1$International Centre for Radio Astronomy Research, Curtin University, Bentley, WA 6102, Australia\\
}%

\date{Accepted XXX. Received YYY; in original form ZZZ}

\pubyear{2018}

\begin{document}
\label{firstpage}
\pagerange{\pageref{firstpage}--\pageref{lastpage}}
\maketitle

\begin{abstract}
A method is presented to limit the volumetric density of repeating fast radio bursts based on the number (or lack) of repeating bursts identified in a survey. The method incorporates the instantaneous sensitivity of the instrument, its beam pattern, and the dwell time per pointing, as well as the energy and timing distribution of repeat bursts. Applied to the Australian Square Kilometre Array Pathfinder's (ASKAP's) Commensal Real-time ASKAP Fast Transients (CRAFT) `lat50' survey, the presence of an FRB similar to FRB~121102 is excluded within a volume of $1.9 \cdot 10^6$\,Mpc$^3$ at 95\% confidence level (C.L.). Assuming a burst energy cut-off at $10^{42}$\,erg, the 95\% C.L.\ upper limit on the population density of repeating FRBs in the current epoch is $27$\,Gpc$^{-3}$, assuming isotropic (unbeamed) emission. This number is much lower than expected from even rare scenarios such as magnetar formation in gamma-ray bursts.
Furthermore, the maximally allowed population under-predicts the observed number of single bursts in the survey. Comparisons with the observed dispersion measure distribution favours a larger population of less rapidly repeating objects, or the existence of a second population of non-repeating FRBs. In any scenario, FRB~121102 must be an atypical object.
\end{abstract}

\begin{keywords}
methods: analytical -- methods: statistical -- radio continuum: transients
\end{keywords}

\section{Introduction}
\label{sec:intro}

Fast Radio Bursts (FRBs) --- millisecond-duration extragalactic bursts of radio waves --- are currently one of the most enigmatic astronomical phenomena. Two fundamental questions regarding their nature are: what are the source(s) of FRBs, and do all FRBs repeat? As of writing, two FRBs --- FRB~121102 \citep{2014ApJ...790..101S,2016Natur.531..202S}, and FRB~180814.J0422+73 \citep{2019Natur.566..230C,2019Natur.566..235C} --- are known to repeat, while numerous follow-up observations (e.g.\ \citet{2015ApJ...799L...5R}) have failed to find secondary bursts from the 60--70 other known FRBs \citep{2016PASA...33...45P}. This is in contrast to the numerous theories of FRB origin, which, while perhaps now less numerous than the number of FRBs themselves, continues to grow \citep{2018NatAs...2..842P}.

The host galaxy of FRB~121102 is located at $z=0.19273$ \citep{2017Natur.541...58C}, with the dispersion measure (DM) of FRB~180814.J0422+73 implying it lies at $z \lesssim 0.1$ \citep{2019Natur.566..235C}. The majority of the remaining FRB population exhibits a DM--fluence relation consistent with a cosmological population extending beyond a redshift of 1 \citep{2018Natur.562..386S}. It may be that apparently once-off FRBs are simply the most powerful bursts from intrinsically repeating objects located at larger distances. In this case, cataclysmic progenitor scenarios --- typically due to the merger of compact objects --- would be excluded.

The other clear possibility is that FRBs may belong to two or more distinct source classes (see e.g.\ \citet{2018NatAs...2..839C}). For instance, cataclysmic events may explain the majority of observed FRBs --- which are intrinsically brighter than the bursts observed from FRB~121102 \citep{2017ApJ...850...76L,2018Natur.562..386S} --- while a non-cataclysmic scenario may explain repeating FRBs.  The historical precedent is the case for gamma-ray bursts (GRBs), which are now understood to belong to at least two different classes \citep{1993ApJ...413L.101K}. Indeed, soft gamma repeaters initially formed part of the GRB population, until identified as repeating objects in the local universe.

One method of identifying the source of FRBs --- repeaters or otherwise --- is to compare the population distribution of FRBs with those of hypothesised FRB progenitors. Clearly, for any candidate class to be a plausible FRB progenitor, its volumetric rate/density must be at least as great as the observed rate/density of FRBs. If the population rate/density of a hypothesised progenitor is greater than the FRB density, then the hypothesis is plausible, unless the difference is so great that additional physics must be invoked to explain why such a small fraction of that population produces FRBs.

Limits have only very recently been placed on the population of FRBs. \citet{2018MNRAS.481.2320L} use observations of 33 FRBs from several instruments to fit the normalised FRB luminosity function. They fit a Schechter luminosity function, finding a power-law distribution of burst strengths with cumulative count index $\gamma=-1.56_{-0.20}^{+0.21}$ and a cut-off luminosity of $2 \cdot 10^{44}$\,erg/s, corresponding to a burst energy of $2 \cdot 10^{41}$\,erg for a characteristic burst duration of 1\,ms. No estimate is made of the absolute rate.

\citet{2018arXiv181109483D} take a sample of 17 FRBs detected by the Parkes radio telescope. The authors find a break in the burst energy distribution at $(3.7 \pm 0.7) \cdot 10^{40}$\,erg, and assuming once-off bursts, find an FRB rate in the current epoch ($z=0$) of $(1.8 \pm 0.3) \cdot 10^4$\,Gpc$^{-3}$\,yr$^{-1}$. The former result is consistent with the break found by \citet{2019PASA...36....9J}, but its ultimate cause is quite possibly due to detection biases \citep{2018MNRAS.474.1900M} against FRBs with signal-to-noise below $16 \sigma$ in the Parkes sample. The latter result is calculated while ignoring the effects of beamshape \citep{2018MNRAS.474.1900M,2019PASA...36....9J}, and thus their results can be quantitatively discounted.

\citet{2019MNRAS.484.5500C} use a Monte Carlo simulation to explicitly consider a population of repeating FRBs, and estimate the results of FRB follow-up observations by a range of telescopes, and an all-sky survey by Parkes. The authors are unable to rule out a single population of repeating FRBs with properties similar to FRB~121102. For two particular FRBs however, the power-law index of the differential burst energy distribution is constrained to be between $-1$ and $-2$.

In this contribution, I develop a method to place limits on the volumetric density of repeating FRBs only, as a function of their properties. The motivation is the lack of repeating bursts detected during the Australian Square Kilometre Array Pathfinder's (ASKAP's) Commensal Real-time ASKAP Fast Transients (CRAFT) `lat50' survey of fields at Galactic latitudes of $|b| = 50^{\circ} \pm 5^{\circ}$ \citep{2018Natur.562..386S}. The advantage of this survey is the wide field of view, and the long observation times spent observing single fields, which is ideal for searching for repeat bursts.

The other reason for focussing on repeating FRBs is to remove dependencies on population modelling: a single strong repeating FRB --- such as FRB~121102 --- cannot be mimicked by multiple weaker FRBs. The downside of focussing on repeating FRBs is that any constraints will only ever apply to FRBs repeating above some rate, since there is no way of constraining arbitrarily rare repeat behaviour.

A method to calculate the volume in which the presence of a repeating FRB with specific properties can be excluded is described in Section~\ref{sec:method}. Using the burst energy distribution of FRB~121102 (discussed in Appendix \ref{sec:repeater}), the method is applied to the ASKAP/CRAFT lat50 survey assuming a Poissonian distribution of burst arrival times in Section~\ref{sec:craft}. Section~\ref{sec:weibull} extends this to the case of a Weibull distribution of burst arrival times, while Section~\ref{sec:varying} examines the effects of changing FRB repeater properties on the exclusion volume.

Section~\ref{sec:volume} extends the methods of Section~\ref{sec:method} to include volumes at larger redshifts, where the presence of any single repeating FRB cannot be excluded, but the total population density can be limited. These methods are again applied to the ASKAP/CRAFT lat50 survey, and in particular, comparisons are made to the number and dispersion measure of detected once-off bursts. Results are discussed in Section~\ref{sec:discussion}.

Somewhat unsatisfyingly, the lack of hard predictions for the FRB population density from the numerous progenitor models does not readily allow model tests. However, given that no previous limits of the repeating FRB population density exist, it is hoped that having these first results will motivate theorists to produce such limits in the near future.

\section{Method of limiting FRBs.}
\label{sec:method}

Limits regarding repeating FRBs can be stated in two broad terms:

\begin{itemize}

\item{Limits on a particular FRB repeating more regularly than some rate above a fluence threshold.}

\item{Limits on the population of FRBs with given properties --- including repetition rate --- existing within a volume, e.g.\ within a particular redshift over some region of sky.}

\end{itemize}

The simplest to derive is the first, by observing the location of a known burst. This can be done either with the discovery instrument, or with a more-sensitive follow-up instrument, and respectively are the methods by which the repeating natures of FRB~180814.J0422+73 and FRB~121102 were discovered. However, this yields no information on the properties of FRBs in general.

The second kind of statement, which is the most useful, is also the most complex to derive. In general, it requires modelling both repeating and once-off FRB populations, their redshift distributions, and probabilities of any given FRB having a specific set of properties. The general ambiguity in fitting the number of bursts observed in an FRB survey is between a population of rare but powerful FRBs, and a more numerous population of less powerful --- or more distant --- objects. This is further complicated by uncertainties in beamshape and the slope of the source-count distribution to systematically affect a survey's sensitivity to a given model \citep{2018MNRAS.474.1900M}.

The method used here to simplify this problem is to only consider the number --- or lack -- of repeating FRBs found in a survey. By calculating the probability of observing multiple bursts from a repeating FRB under some assumption about its properties, and setting this probability to an appropriate confidence limit (e.g.\ 95\%), the number of such FRBs within a volume probed by a survey can be limited. A clear disadvantage to this approach is that it cannot limit the population distribution of once-off FRBs. However, such a limit on repeating FRBs will be independent of the properties of the rest of the FRB population.\footnote{When FRB discoveries become so numerous that the chance detection of two independent FRBs with the same dispersion measure from the same direction is non-negligible, this will no longer apply.} Another advantage of this approach is that the probability of two or more detections scales with the square of sensitivity, which both reduces the importance of sidelobes and their systematic effects, and reduces the influence of FRBs arriving from the distant universe.

The numerous observations of FRB~121102 also define an archetypical source in a way which is impossible for singly detected FRBs. Limits on the number of FRB~121102-like objects in the universe therefore touch on the key question: do all FRBs repeat?

\subsection{Properties of a repeating FRB}

The full range of properties that could describe a repeating FRB is exceptionally broad, including burst rate, time distribution, energy, duration, frequency structure, etc. Here, repeating FRBs are primarily characterised via the cumulative burst rate $R$ above a burst energy threshold $E$:
\begin{eqnarray}
R(E>E_0) & = &  R_0 \left( \frac{E}{E_0} \right)^{\gamma}, \label{eq:intrinsic_rate}
\end{eqnarray}
where $\gamma$ is the power-law index. Such a distribution has been observed for the first known repeating burst source, FRB~121102 \citep{2017ApJ...850...76L}. In Appendix~\ref{sec:repeater}, default values of $R_0$, $E_0$, and $\gamma$ are derived, finding:
\begin{eqnarray}
R_0 & = & 7.4_{-4.8}^{+4.0}\,{\rm day}^{-1} \label{eq:defaults} \\
E_0 & = & 1.7 \cdot 10^{38}\,{\rm erg}  \nonumber \\
\gamma & = & -0.9 \pm 0.2. \nonumber
\end{eqnarray}
These are broadly consistent with the model of intrinsic properties presented by \citet{2017ApJ...850...76L}. No minimum of cut-off energy is assumed or required, while a maximum energy cutoff will not be required until Section~\ref{sec:volume}. Further properties discussed in Appendix~\ref{sec:repeater} which will soon become relevant are the burst bandwidth, $\Delta \nu_{\rm FRB}$, taken here to be 420\,MHz; and a typical burst duration, $\Delta t_{\rm FRB}$, of 0.2--2\,ms.

In general, the analysis presented below is adaptable to any distribution of $R$, with a model including a burst energy cutoff at $E=E_{\rm cut}$ analysed in Section~\ref{sec:volume}. For the remainder of this section, the simple power-law model above is retained, due to its analytic simplicity.

\subsection{Probability of observing multiple bursts}
\label{sec:pmult}

For bursts with independent (Poissonian) arrival times, an expected number of $2.36$ bursts is required for a $1\,\sigma$ ($68$\%) chance of detecting two or more bursts. Therefore, the lack of repeat bursts allows an exclusion limit, $\lambda_{\rm lim}$, of $2.36$ or more bursts to be set at 68\% confidence level (C.L.). Similarly, $90$, $95$, and $99.7$\% C.L.s correspond to $\lambda_{\rm lim}$ of $3.89$, $4.84$, and $7.83$ respectively. These numbers apply only for the Poissonian case. In general however, for any given model of the time-distribution of bursts, there will exist some expected number of events $\lambda_{\rm lim}$ at which two or more bursts would be detected with some confidence level (C.L.). The Poissonian case is considered in Section~\ref{sec:craft}, a non-Poissonian case in Section~\ref{sec:weibull}, and for now, $\lambda_{\rm lim}$ is left as a free parameter.

Given $\lambda_{\rm lim}$, and an observation time $T_{\rm obs}$ spent on-source, the presence of a repeating FRB with intrinsic rate $R_{\rm lim}$ can be excluded provided:
\begin{eqnarray}
R_{\rm lim} & = & (1+z) \frac{\lambda_{\rm lim}}{T_{\rm obs}},
\end{eqnarray}
where the factor of $(1+z)$ is due to the time-dilation effects of redshift.

The value of $E$, $E_{\rm lim}$, above which the rate is $R_{\rm lim}$, can be found by inverting equation~(\ref{eq:intrinsic_rate}):
\begin{eqnarray}
E_{\rm lim} & = & E_0 \left(\frac{R_{\rm lim}}{R_0}\right)^{\frac{1}{\gamma}} \nonumber \\
& = & E_0 (1+z)^{\frac{1}{\gamma}} \left(\frac{\lambda_{\rm lim}}{T_{\rm obs} R_0}\right)^{\frac{1}{\gamma}}. \label{eq:elim}
\end{eqnarray}
Intrinsic pulses of strength $E_{\rm lim}$ must then be observable, this being a function of $F$ and $z$. Given a fluence threshold $F_{\rm th}$, $z_{\rm lim}$ is defined as the redshift at which pulses of intrinsic strength $E_{\rm lim}$ are just observable at threshold $F_{\rm th}$.

The fluence $F$ at which a transient with total energy $E$ arrives is given by \citet{1996ASPC...88..107M}, and discussed in the context of FRBs by \citet{2018MNRAS.tmp.1976M}:
\begin{eqnarray}
F(\nu) & = & \frac{(1+z)^{2+\alpha}}{\Delta \nu_{\rm FRB}} \frac{E}{4 \pi D_L^2}. \label{eq:Fnu}
\end{eqnarray}
Here, $D_L$ is the luminosity distance to the source, $\Delta \nu_{\rm FRB}$ is the intrinsic bandwidth of the bursts, and $\alpha$ is the spectral index ($F(\nu) \propto \nu^{\alpha}$, i.e.\ the k-correction is given by $(1+z)^{\alpha}$). The other factor of two in the exponent of $1+z$ accounts for both bandwidth compression, and that time-dilation of the pulse duration does not affect the fluence as it does luminosity. 

Equation~(\ref{eq:Fnu}) applies if the total burst width in frequency space is much greater than the observing bandwidth, $\Delta \nu_{\rm obs}$. However, most bursts from FRB~121102 appear to have relatively narrow and complex frequency-domain structure \citep{2019ApJ...876L..23H}, where the use of a spectral index and conventional k-correction does not apply.

In order to account for this, observe that there will be a critical value of redshift, $z_{\rm crit}$, at which the burst bandwidth matches the observation bandwidth, $\Delta \nu_{\rm obs}$:
\begin{eqnarray}
z_{\rm crit} & = & \frac{\Delta \nu_{\rm FRB}}{\Delta \nu_{\rm obs}} -1. \label{eq:zcrit}
\end{eqnarray} 
For $z<z_{\rm crit}$, only a fraction of the burst will be contained in the observation bandwidth, in which case the experimental fluence threshold can be compared directly to the calculated FRB fluence. Therefore equation~(\ref{eq:Fnu}) is used to calculate the observed fluence $F_{\rm obs}$ with $\alpha=0$, i.e.:
\begin{eqnarray}
F_{\rm obs} & = & \frac{(1+z)^{2}}{\Delta \nu_{\rm FRB}} \frac{E}{4 \pi D_L^2} ~~~[z \le z_{\rm crit}]. \label{eq:Fobsnear}
\end{eqnarray}
For $z>z_{\rm crit}$, the total burst energy will be contained within only a fraction of the detection band. The observed fluence when averaged over the detection band, $F_{\rm obs}$, is given by:
\begin{eqnarray}
F_{\rm obs} & = &  \frac{(1+z)}{\Delta \nu_{\rm obs}} \frac{E}{4 \pi D_L^2} ~~~[z > z_{\rm crit}].
\label{eq:Fobsfar}
\end{eqnarray}
Setting the observed fluence equal to the experimental fluence threshold $F_{\rm th}$, and equating this with $E_{\rm lim}$ using equations~(\ref{eq:Fobsnear}) and (\ref{eq:Fobsfar}), gives a solution for $z_{\rm lim}$ for any given value of $\lambda$ and, hence, confidence level:
\begin{eqnarray}
\frac{ D_L(z_{\rm lim})^2}{(1+z_{\rm lim})^{2+\frac{1}{\gamma}}} ~ = ~ \frac{E_0}{4 \pi F_{\rm th} \Delta \nu_{\rm FRB}} \left(\frac{\lambda_{\rm lim}}{T_{\rm obs} R_0}\right)^{\frac{1}{\gamma}} ~[z_{\rm lim} \le z_{\rm crit}] \nonumber \\
\frac{ D_L(z_{\rm lim})^2}{(1+z_{\rm lim})^{1+\frac{1}{\gamma}}} ~ = ~ \frac{E_0}{4 \pi F_{\rm th}  \Delta \nu_{\rm obs}} \left(\frac{\lambda_{\rm lim}}{T_{\rm obs} R_0}\right)^{\frac{1}{\gamma}} ~[z_{\rm lim} > z_{\rm crit}]. \label{eq:zlim}
\end{eqnarray}
While equation~(\ref{eq:zlim}) must be solved numerically, the dependence of the left hand side is mostly quadratic in $z$ in the nearby universe, and numerical convergence is rapid. Either regime in $z_{\rm lim}$ can be tested-for first, with a resulting $z_{\rm lim}$ in the incorrect range indicating that the other equation must be used.

\subsection{Limits as a function of solid angle}
\label{sec:solid_angle}

The fluence threshold $F_{\rm th}$ above which an experiment detects a burst is subject to several factors, such as burst duration and frequency dependence. The primary effect however, as discussed in \citet{2018MNRAS.474.1900M}, is location in the telescope beam. Given a nominal experimental threshold $F_0$ at beam centre, the effective threshold in any given direction is given by:
\begin{eqnarray}
F_{\rm th} & = & \frac{F_0}{B}, \label{eq:fth_beam}
\end{eqnarray}
where $B$ is the value of the beam power pattern at that position ($B=1$ at beam centre).

For a single survey field, different fractions of the sky will be covered by different thresholds $F_{\rm th}$, and hence a different $z_{\rm lim}$ applies at each position. The total solid angle $\Omega(z_{\rm lim})$ at which a limit of $z_{\rm lim}$ applies can be found using the `inverse beam pattern', $\Omega(B)$, as defined by \citet{2019PASA...36....9J}, which gives the total solid angle at which any given value of $B$ applies. The solid angle $\Omega(z_{\rm lim})$ is then given by:
\begin{eqnarray}
\Omega(z_{\rm lim}) dz_{\rm lim} & = & \Omega(B(z_{\rm lim},T_{\rm obs})) dB \frac{dz_{\rm lim}}{dB}, \label{eq:single_omega_zlim}
\end{eqnarray}
where $T_{\rm obs}$ is the total time spent observing that field, and $B(z_{\rm lim},T_{\rm obs})$ is the required value of beam sensitivity to limit $z<z_{\rm lim}$ in observation time $T_{\rm obs}$. It, and its derivative with respect to $z$, can be calculated from equations~(\ref{eq:zlim}) and (\ref{eq:fth_beam}). Since $\Omega(B)$ is usually calculated numerically as a histogram in $B$ however, it is practically much easier to produce $\Omega(z_{\rm lim})$ as a histogram as well, in which case the factors $dz_{\rm lim}$ and $dB$ become histogram bin widths, and $dB/dz_{\rm lim}$ is the mapping between bins. This procedure tends to work very well for log-spaced binning in $B$ and $z_{\rm lim}$.

For a survey over $i=1\ldots N$ fields with observation time $T_i$ per field, the total solid angle at which a limit of $z_{\rm lim}$ applies is simply the sum of equation~(\ref{eq:single_omega_zlim}) over all fields:
\begin{eqnarray}
\Omega(z_{\rm lim}) dz & = & \sum_{i=1}^{N_f} \Omega(B(z_{\rm lim},T_i)) dB \frac{dB}{dz}. \label{eq:omega_zlim}
\end{eqnarray}

\subsection{Total limited volume}

Since $z_{\rm lim}$ gives the value of $z$ \emph{within} which the presence of a repeating FRB can be excluded, the total solid angle $\Omega_{\rm lim}(z)$ over which the presence of an FRB at redshift $z$ is excluded can be calculated as:
\begin{eqnarray}
\Omega_{\rm lim}(z) & = & \int_{z_{\rm lim}=z}^{\infty} \Omega(z_{\rm lim}) d z_{\rm lim}. \label{eq:omega_lim_z}
\end{eqnarray}
The total volume over which the presence of a repeating FRB can be excluded, $V_{\rm lim}$, can then be calculated as:
\begin{eqnarray}
V_{\rm lim} & = & \int_0^{\infty} \Omega_{\rm lim}(z) D_H \frac{(1+z)^2 D_A^2}{E(z)} dz \label{eq:vlim}
\end{eqnarray}
where the terms in the integrand represent the comoving volume element, $D_A$ is the angular diameter distance, $D_H$ the Hubble distance, and $E(z)$ scales the Hubble parameter $H(z)$ from its present value. Here, a minimal $\Lambda$CDM cosmology appropriate to $z \lesssim 1$ is used:
\begin{eqnarray}
H(z) & = & H_0 E(z) \\
E(z) & = & \sqrt{\Omega_m (1+z)^3 + \Omega_{\Lambda}}, \label{eq:cosmology}
\end{eqnarray}
with $H_0=67.74$\,km\,s$^{-1}$\,Mpc$^{-1}$, matter density $\Omega_m=0.31$, and dark energy density $\Omega_{\Lambda}=0.69$ \citep{2016A&A...594A..13P}.

\subsection{Effects of overlapping fields}
\label{sec:procedure}

The formulae above assume a single mapping between beam sensitivity $B$ and pointing direction. However, surveys attempting to fully cover a region of sky will define survey fields such that beams from neighbouring fields partially overlap, e.g.\ at the half-power point. In general, any given point of sky will be observed by $j=1 \ldots M$ fields at beam sensitivity $B_j$ over a duration $T_j$.

In order to simplify the problem, note that equation~(\ref{eq:zlim}) contains a factor of $(T_{\rm obs} F_{\rm th}^{\gamma})^{-1/\gamma}$, where $T_{\rm obs} F_{\rm th}^{\gamma}$  represents the effective exposure to a power-law distribution of fluences. For a survey with $i=1 \ldots N_p$ pointings at beam values $B_i$ (so that $F_{\rm th}=F_0/B$) for times $T_i$, the necessary replacement is:
\begin{eqnarray}
T_{\rm obs} F_{\rm th}^{\gamma} & \to & \sum_{i=1}^{N_p} T_i \left(\frac{F_0}{B_i}\right)^{\gamma}  \label{eq:Feff_factor}
\end{eqnarray}
for each direction in the sky. This can then be inserted into the right-hand side of equation~(\ref{eq:zlim}), giving (e.g.\ for the $z<z_{\rm crit}$ regime):
\begin{eqnarray}
\frac{ D_L(z_{\rm lim})^2}{(1+z_{\rm lim})^{2+\frac{1}{\gamma}}} & = & \frac{E_0}{4 \pi \Delta \nu_{\rm FRB} F_{0}} 
 \left( \frac{\lambda_{\rm lim}}{R_0 \left( \sum_{i=1}^{N_p} \frac{T_i}{B_i^\gamma}\right)} \right)^{\frac{1}{\gamma}}. \label{eq:zeff_factor}
\end{eqnarray}
This then allows a value of $z_{\rm lim}$ to be found for each position on sky, and the calculation of $\Omega(z_{\rm lim})$ in equation~(\ref{eq:single_omega_zlim}) becomes a sum over all positions. Note however that condensing multiple observations into a single figure of merit is only possible if burst arrival times are uncorrelated, i.e.\ they follow a Poissonian distribution.

The calculation of $z_{\rm lim}$ for each and every position on the sky is relatively complicated however. Furthermore, most directions will be dominated by the contribution from a single survey field, except small regions at the overlap points between adjacent survey fields. The following procedure is therefore recommended. In order to calculate the total solid angle $\Omega_{\rm lim}(z)$ at which a repeating FRB can be excluded, use a truncated beam pattern, by setting $B=0$ at regions of the beam pattern covered at greater sensitivity by a different survey field (e.g.\ beyond the half-power points). This will reduce $\Omega(B)$ at low values of $B$, and limit $\Omega_{\rm lim}(z)$ in the low-z region to the actual survey area.

However, when calculating volumetric limits, use the full beam pattern. This will still underestimate $V_{\rm lim}$, but not by as much as using a truncated beam pattern. To illustrate this, consider the probability of observing two or more bursts from a region of sky nominally inside one survey field, but with some sensitivity from an outer beam sidelobe in a secondary survey field. Clearly, this probability is given by the probability of seeing two or more bursts from the first field, plus the probability of seeing two or more bursts from the second field, plus the probability of seeing exactly one burst from each field. Truncating the beamshapes accounts for only the first term, while using untruncated beamshapes accounts for the first two. Incorporating the third requires the calculations given in equations~(\ref{eq:Feff_factor}) and (\ref{eq:zeff_factor}) above.

\section{Limits from the CRAFT `lat50' survey}
\label{sec:craft}

In this section, all results and associated confidence levels (C.L.) correspond to those for a Poissonian distribution of pulse arrival times, with $\lambda_{\rm lim}$ given by the second column in Table~\ref{tbl:weibull}. The non-Poissonian arrival time distribution of FRB~121102 will be analysed in Section~\ref{sec:weibull}.

The first results of the Commensal Real-time ASKAP Fast Transients (CRAFT; \citet{2010PASA...27..272M}) project using the Australian Square Kilometre Array Pathfinder (ASKAP; \citet{2009IEEEP..97.1507D}) are described in \citet{2018Natur.562..386S}. Using a total of approximately $1326$ antenna-days in flye's eye mode, the initial observations surveyed fields at galactic latitudes $|b| = 50^{\circ} \pm 5^{\circ}$. These `lat50' observations covered a frequency range of $1128$ to $1464$\,MHz, with frequency and time resolutions of $1$\,MHz and $1.2656$\,ms respectively. A total of $20$ FRBs were discovered. Importantly, each field was revisited for many antenna days, and \citet{2018Natur.562..386S} use the non-observation of repeating pulses to show that the observed properties of these FRBs are inconsistent with those of FRB~121102. However, this does not exclude that these pulses may be rare and bright emission from distant repeating sources.

\begin{figure}
\begin{center}
\includegraphics[width=\columnwidth]{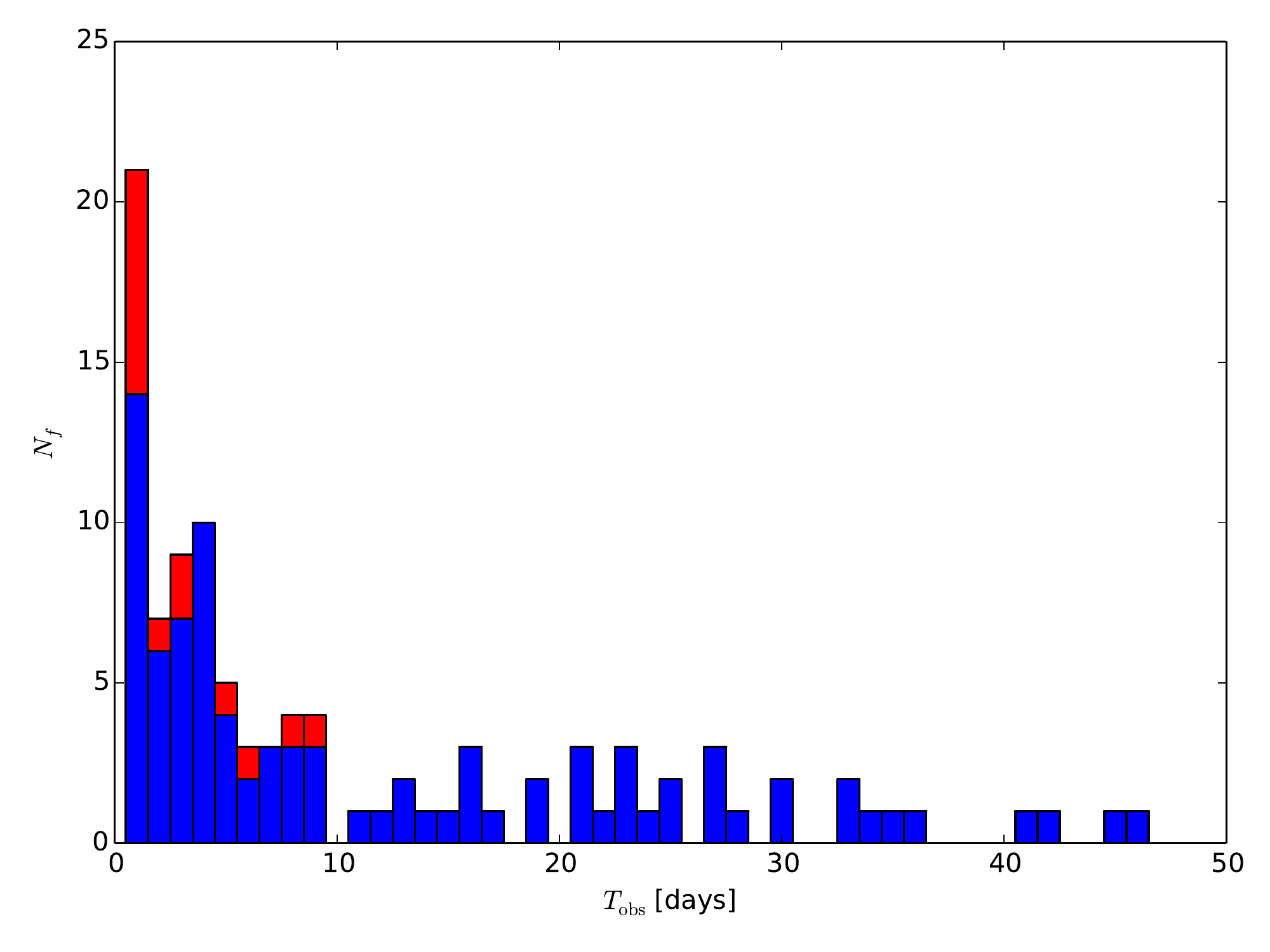}
\caption{
Number of fields $N_f$ with observation time per field $T_{\rm obs}$ for the ASKAP/CRAFT `lat50' survey reported in \citet{2018Natur.562..386S}. Primary fields at Galactic latitude $|b|=\pm50^{\circ}$ are shown in blue; secondary fields are shown in red.
} \label{fig:craft_tobs}
\end{center}
\end{figure}

Figure~\ref{fig:craft_tobs} shows the time on-sky for each of the pointings in the lat50 survey. These have been adjusted to the total effective observation time, after accounting for efficiency losses, of $1108.9$~days \citep{2019PASA...36....9J}.

A total of $57$ primary fields were studied at $|b|=\pm50^{\circ}$, separated by $5.4^{\circ}$ in Galactic longitude (shown in blue). For each field, the total time observation time was spread between January 2017 and March 2018. Coverage was not uniform, since fields with high elevation angles were preferentially targeted, and observations times with antennas undergoing commissioning were opportunistic. Additionally, a small number of other fields were included in the survey. Most of these varied Galactic latitude by $\pm5^{\circ}$ to maximise overlap with the beam of the Murchison Widefield Array (MWA) \citep{2018ApJ...867L..12S}, while some targeted high-latitude FRBs detected by the Parkes radio telescope. The durations of these pointings are indicated in red in Figure~\ref{fig:craft_tobs}.

\subsection{ASKAP/CRAFT lat50 beamshape}

\begin{figure}
\begin{center}
\includegraphics[width=\columnwidth]{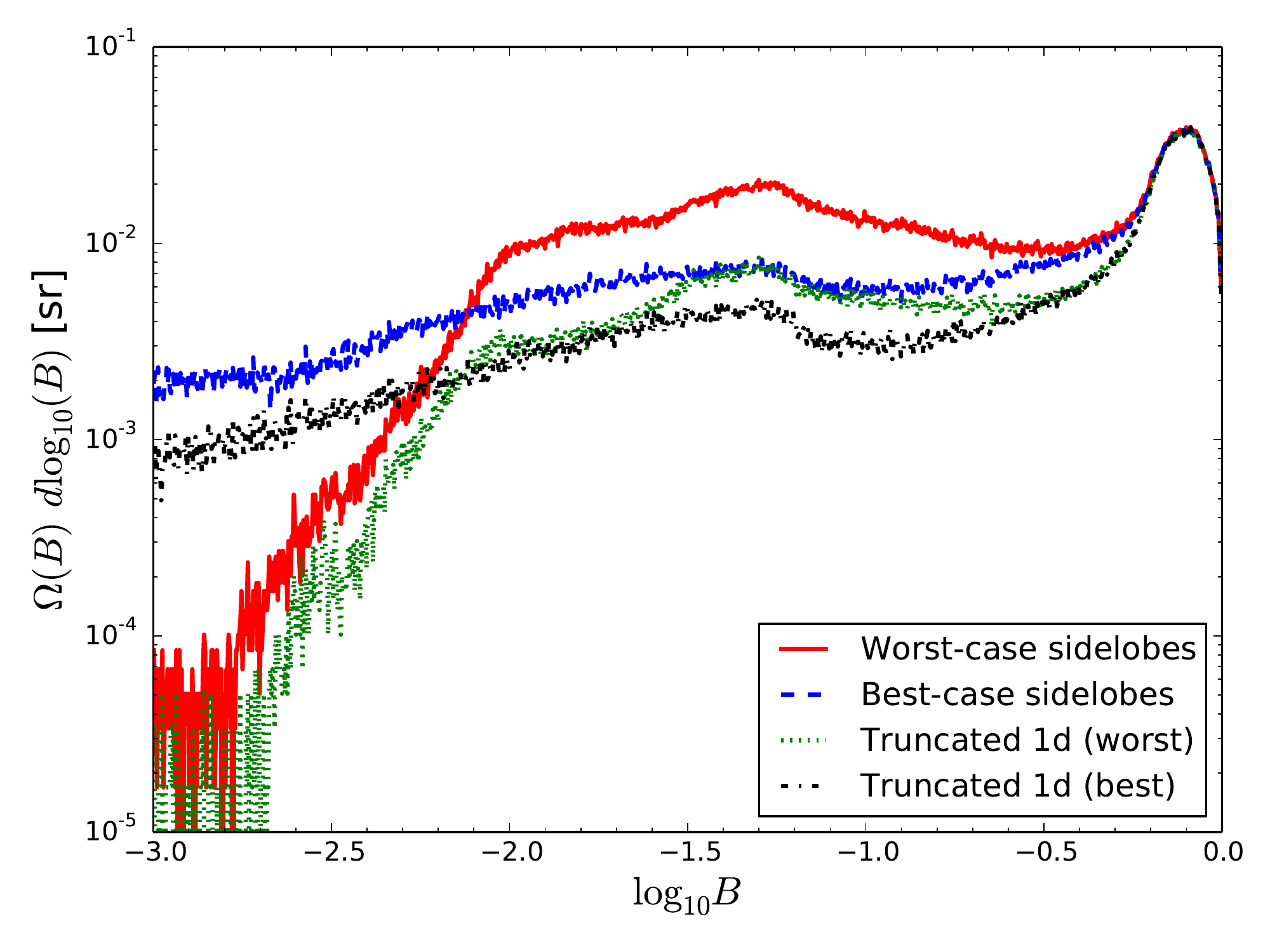}
\caption{
Inverse beamshapes $\Omega(B)$ for the closepack36 beam configuration. Best and worst-case sidelobes correspond to different assumptions made during beam calibration --- see \citet{2019PASA...36....9J} for details. `Truncated 1d' results are calculated by first setting beamshapes $B(\Omega)$ to zero when fields overlap in Galactic longitude (i.e.\ in one dimension of the beamshape).
} \label{fig:omega_B_askap}
\end{center}
\end{figure}

The beam configuration of ASKAP for most of the survey was `closepack36', with total sensitivity $\Omega(B)$ derived in \citet{2019PASA...36....9J}. The calibration procedure resulted in uncertainties regarding the outer beam sidelobes --- $\Omega(B)$ is plotted for both the `worst case' (largest sidelobe) and `best case' (lowest sidelobe) beamshapes in Figure~\ref{fig:omega_B_askap}.

Following the procedure suggested in Section~\ref{sec:procedure}, $\Omega(B)$ is also calculated by truncating the beams in the direction of Galactic longitude (i.e.\ one dimension), where adjacent fields overlapped.

For all beamshapes, the peak in the range $-0.3 < \log_{10}(B) < 0$ represents the solid angle spanned by the 36 ASKAP beams near FWHM, while $\Omega(B)$ for lower beam power $B$ is the solid angle spanned by sidelobes. The downturns below $\log_{10}(B) < -2$ for the worst-case beams is artificial, and due to the beamshapes being calculated over an $8^{\circ} \times 8^{\circ}$ region only.\footnote{The integral of $\Omega(B)$ over all $B$ must come to $4 \pi$\,sr.}

Since the goal of this work is to calculate upper limits from the non-observation of repeat pulses, from now on only the best-case sidelobes are considered, which (despite the nomenclature) will produce the worst limits on the presence of FRBs through lower total sky coverage.

\subsection{ASKAP/CRAFT lat50 sensitivity}
\label{sec:askap_sensitivity}

The antenna-average fluence threshold of ASKAP FRB observations is approximately $26$\,Jy\,ms \citep{2018Natur.562..386S,2019PASA...36....9J} for a pulse contained entirely within the sampling time of 1.2656\,{\rm ms}. The in-channel smearing due to the $1$\,MHz channel width is approximately equal to the sampling time resolution for a DM of $262$\,pc\,cm$^{-3}$ at band centre (1.296\,GHz). The width of the FRB itself, $\Delta t_{\rm FRB}$, will also increase the threshold by smearing the burst over more system noise. This produces a DM-dependent fluence threshold $F_0({\rm DM},\Delta t_{\rm FRB})$ approximately equal to:
\begin{eqnarray}
F_0 & = & 26\,{\rm Jy\,ms}\, \left(1+\frac{\Delta t_{\rm FRB}}{1.2656\,{\rm ms}} + \frac{\rm DM}{262\,{\rm pc\,cm}^{-3}} \right)^{0.5}. \label{eq:F0DM}
\end{eqnarray}
This expression is similar to that used in \citet{2019MNRAS.484.5500C}, but with the sampling time included (and scattering ignored, which \citet{2019MNRAS.484.5500C} ultimately do also).
The dispersion measure (DM) of $557$\,pc\,cm$^{-3}$ for FRB~121102 \citep{2014ApJ...790..101S}, and the burst duration ($\Delta t_{\rm FRB}=0.2$--2\,ms; \citet{2017ApJ...850...76L,2018ApJ...863....2G}), would produce a total time duration of approximately $2.9$--$4.7$\,ms in the ASKAP/CRAFT system (assuming no scattering), and thus thresholds in the range of $47$--$56$\,Jy\,ms.

\begin{figure}
\begin{center}
\includegraphics[width=\columnwidth]{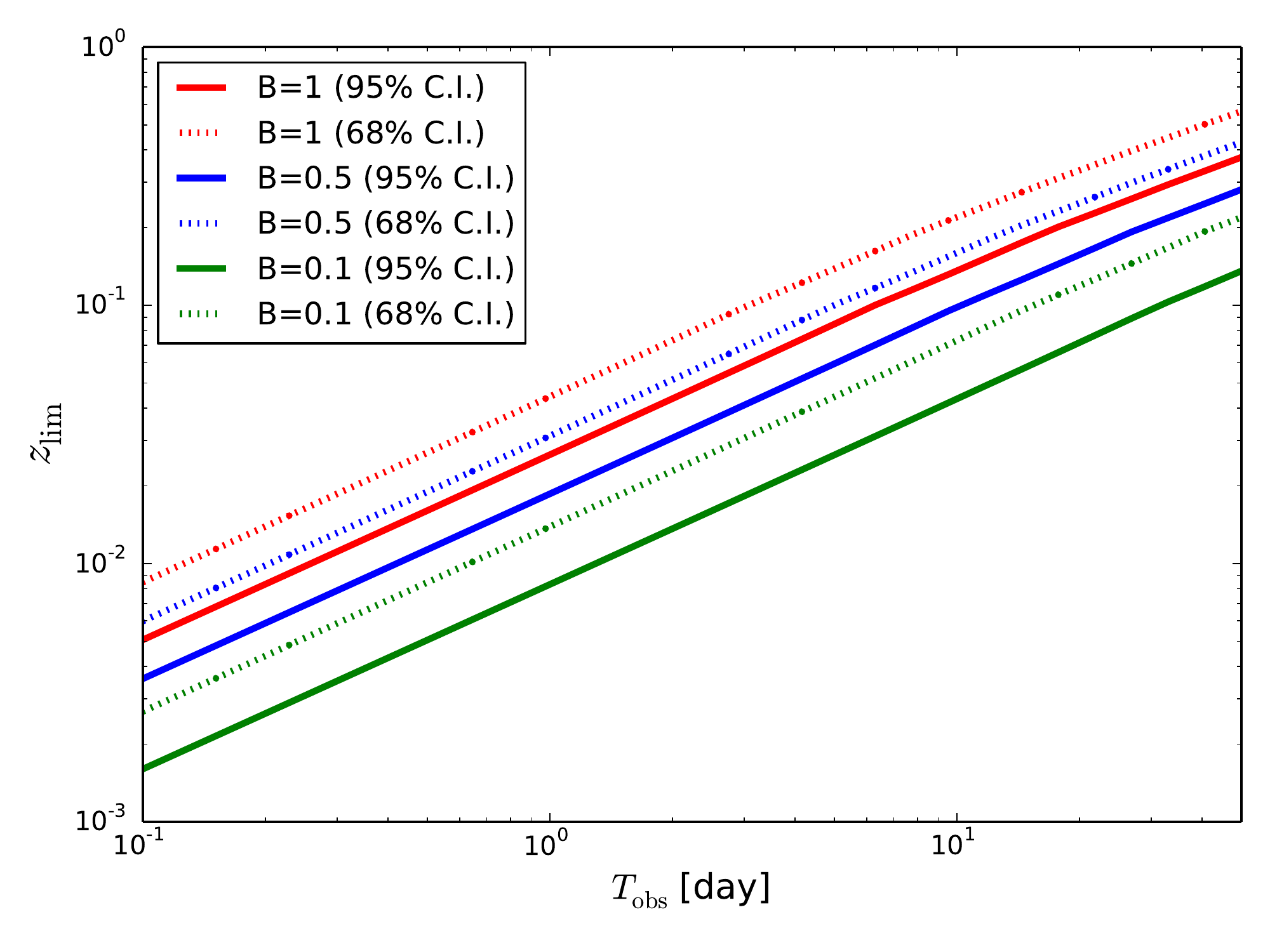}
\caption{
Dependence of $z_{\rm lim}$ on observation time $T_{\rm obs}$ at different beam sensitivities $B$ and confidence levels $C.L.$ for an ASKAP threshold of $F_0=52$\,Jy\,ms, assuming a Poissonian distribution of burst arrival times.
} \label{fig:craft_zlim_t}
\end{center}
\end{figure}

The expected total DM of an FRB can be calculated as per \citet{2004MNRAS.348..999I}; here, a fully ionised, uniform intergalactic medium with an electron number density in the current epoch of $n_e(z=0)=9.898 \cdot 10^{-6} \Omega_b h^2$ and physical baryon density $\Omega_b h^2=0.02264$ \citep{2013ApJS..208...20B} is assumed. No inhomogeneities (see e.g.\ \citet{2014ApJ...780L..33M}) are considered. Doing so (i.e.\ ignoring inhomogeneities) produces an approximate intergalactic DM contribution DM$_{\rm IGM}$ of 100\,pc\,cm$^{-3}$ at $z=0.1$ and 1100\,pc\,cm$^{-3}$ at $z=1$, to which would be added Galactic and host contributions, and the DM from intervening galaxies and halos. For this sample, the total Galactic contribution from the NE2001 model of \citet{2002astro.ph..7156C} is less than $65$ \,pc\,cm$^{-3}$, assuming a Galactic halo contribution of $15$\,pc\,cm$^{-3}$ as per \citet{2018Natur.562..386S}. Thus at $z=1$ the minimum dispersion smearing of a pulse would smear it over four samples. Rather than use a $z$-dependent threshold therefore, the ASKAP fluence threshold $F_0$ is set at twice the nominal value, i.e.\ $F_{0}=52$\,Jy\,ms, to account for burst smearing over four samples. A $z$-dependent threshold is considered in Section~\ref{sec:volume}.

Using this threshold, and noting that $z_{\rm crit} \approx 0.19$ for ASKAP, Figure~\ref{fig:craft_zlim_t} shows $z_{\rm lim}$ corresponding to different values of beam sensitivity $B$ (and hence threshold $F_{\rm th}$) as a function of observation time $T_{\rm obs}$. The most sensitive observations ($B=1$ and $T_{\rm obs}\sim 50$ days) produce $z_{\rm lim}=0.7$ (DM$_{\rm IGM}=770$\,pc\,cm$^{-3}$) at 68\% C.L. This range of $z_{\rm lim}$ is broadly consistent with the assumed threshold, which will only be too optimistic for FRBs near $z=0.7$ with significant host or halo contributions, and too pessimistic for FRBs at low $z$ (corresponding to $z_{\rm lim}$ for low $B$, $T_{\rm obs}$, and 95\% C.L.) with little to no host or halo contribution.

\begin{figure}
\begin{center}
\includegraphics[width=\columnwidth]{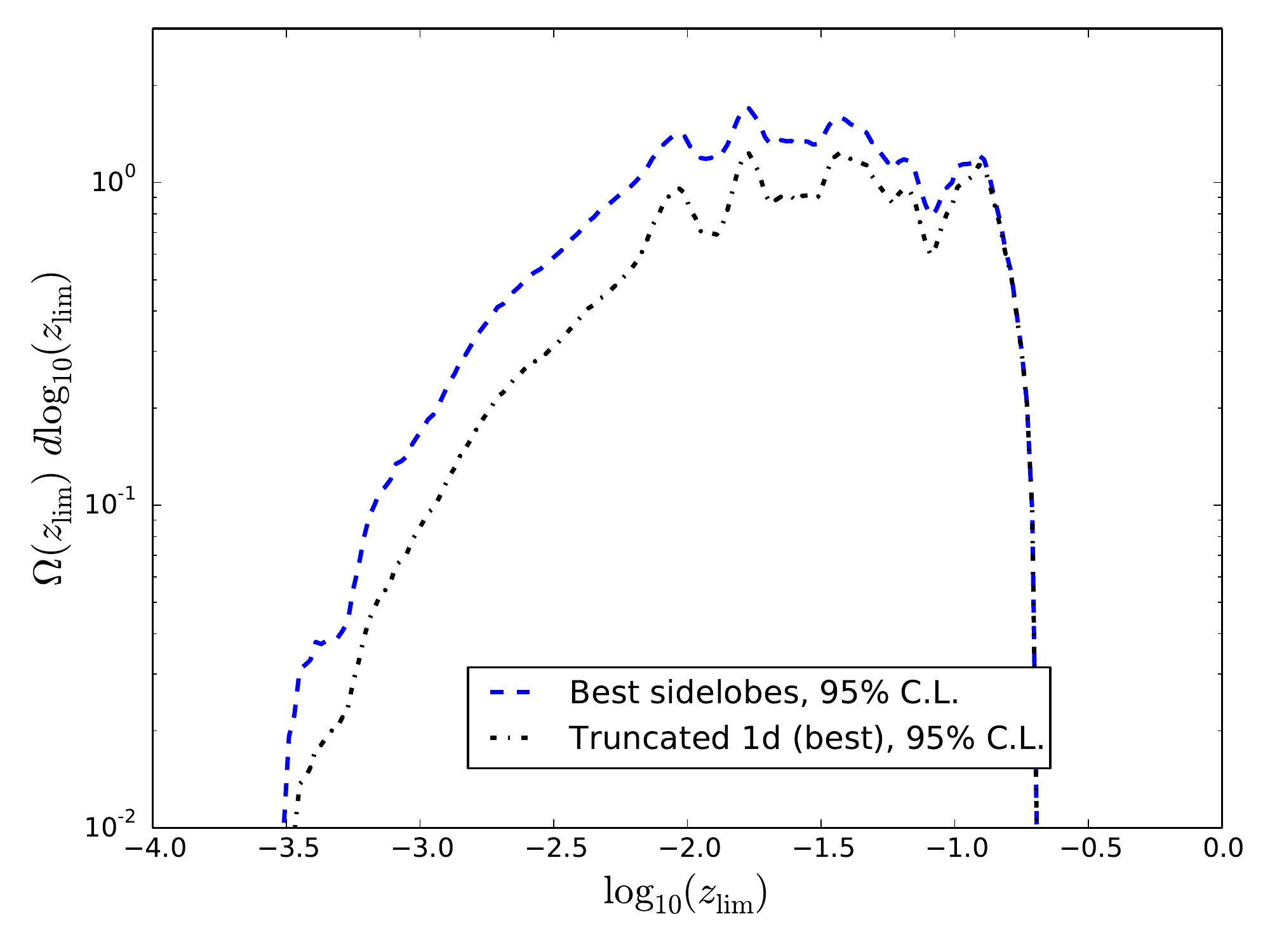}
\caption{
Solid angle $\Omega(z_{\rm lim})$ over which the presence of a repeating FRB with properties given by equations~(\ref{eq:intrinsic_rate}) and (\ref{eq:defaults}) can be excluded within $z<z_{\rm lim}$ at 95\% confidence, assuming a Poissonian distribution of burst arrival times.
} \label{fig:zlim_hist}
\end{center}
\end{figure}

Figure~\ref{fig:zlim_hist} plots $\Omega(z_{\rm lim})$ for the ASKAP/CRAFT lat50 survey, calculated as per equation~(\ref{eq:omega_zlim}). The structure in the range $-2.0 < \log_{10}(z_{\rm lim}) < -0.9$ mimics that of the pointing time histogram. In the truncated case, the removal of sidelobes in one dimension becomes increasingly important at low redshifts. $\Omega(z_{\rm lim})$ is likely underestimated for $\log_{10} (z_{\rm lim}) \le -2.5$, since in this region, far sidelobes --- which are not included in the ASKAP beamshape estimates --- will become important.

\begin{figure}
\begin{center}
\includegraphics[width=\columnwidth]{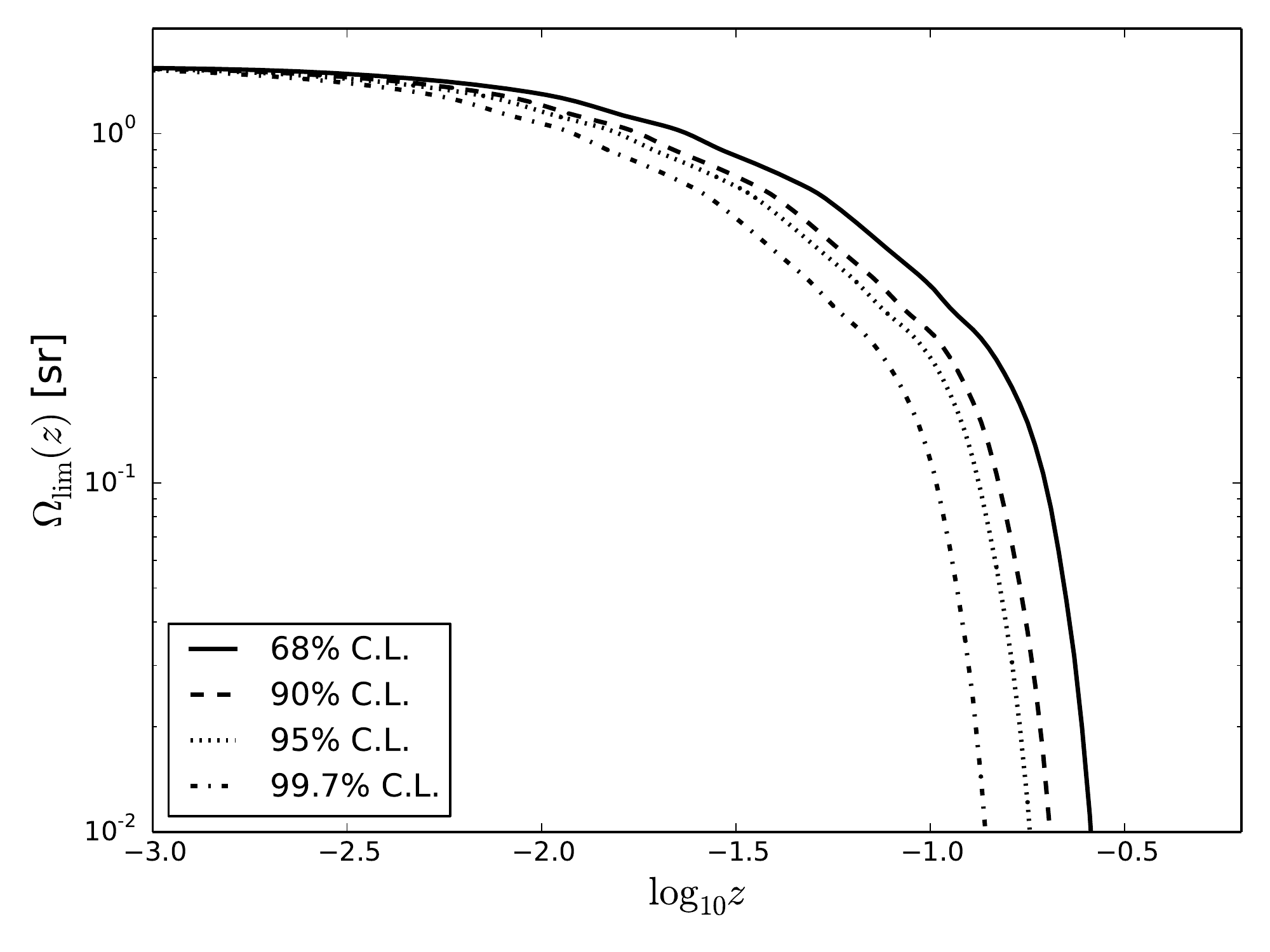}
\caption{
Solid angle $\Omega_{\rm lim}(z)$ over which the presence of a repeating FRB with properties given by equations~(\ref{eq:intrinsic_rate}) and (\ref{eq:defaults}) can be excluded as a function of $z$ at the stated confidence levels (C.L.). Calculations use truncated beams with best-case sidelobes.
} \label{fig:omega_lim_z}
\end{center}
\end{figure}

The region of sky $\Omega_{\rm lim}(z)$ (equation~(\ref{eq:omega_lim_z}) over which the presence of an FRB can be limited is shown in Figure~\ref{fig:omega_lim_z}, for the best-case, truncated beam only (see Section~\ref{sec:procedure}). At low values of $z$, the total solid angle covered converges on $2$\,sr. This is approximately double the survey area which would be calculated using the nominal field of view of $\sim 30$\,deg$^2$ and the total number of survey fields ($N_f=103$), and is due to the influence of sidelobes. Note that the value of $\Omega_{\rm lim}(z)$ as $z \to 0$ is somewhat arbitrary, since any radio telescope will cover all $2 \pi$\,sr of the sky at some non-zero value of $B$. In this case, $\Omega_{\rm lim}(z)$ is only greater than the nominal survey area for $\log_{10}(z)\lesssim -2.0$ at 95\% C.L.

\begin{figure}
\begin{center}
\includegraphics[width=\columnwidth]{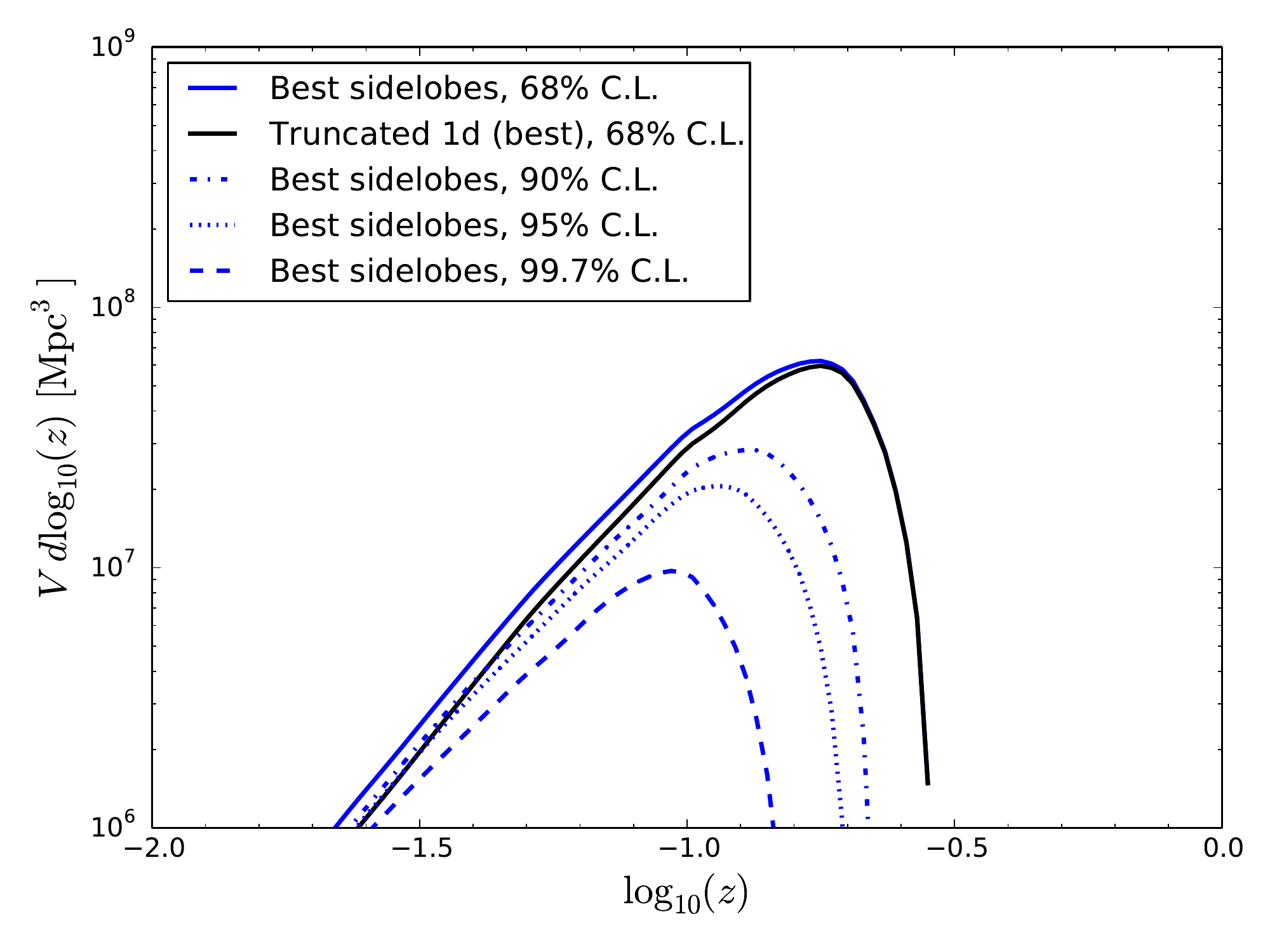}
\caption{
Volumetric limit as a function of $z$ within which the presence of an FRB can be excluded at the stated confidence levels (C.L.). Standard results are for the best-case sidelobes; the truncated beam at $68\%$ C.L. is presented for comparison purposes only. Calculations correspond to the integrand of equation~(\ref{eq:vlim}).  Assumed FRB properties are given by equations~(\ref{eq:intrinsic_rate}) and (\ref{eq:defaults}) with Poissonian arrival times.
} \label{fig:vlim_integrand}
\end{center}
\end{figure}

The influence of low-sensitivity sidelobes on final limits on repetition are shown to be negligible in Figure~\ref{fig:vlim_integrand}, which plots the integrand of equation~(\ref{eq:vlim}). That is, it is a volume-weighted version of the $\Omega_{\rm lim}(z)$ shown in Figure~\ref{fig:omega_lim_z}.
 The much greater volume probed at high $z$ results in a very strong dependence on beam sensitivity, reducing the influence of sidelobes. It also means that the chosen level of confidence has a large effect on the resulting volumetric limits.

\begin{table}
\caption{Volume, $V_{\rm lim}$, within which the presence of a repeating FRB can be limited at the stated level of confidence, C.L. Assumed FRB properties are given by equations~(\ref{eq:intrinsic_rate}) and (\ref{eq:defaults}) with Poissonian arrival times.}\label{tbl:vlim}
\centering
\begin{tabular}{c | c}
\hline
\hline
C.L.\ & $V_{\rm lim}$  \\
 \% & Mpc$^3$ \\
\hline
68	& $2.6 \cdot 10^7$ \\
90	& $1.2 \cdot 10^7$ \\
95	& $8.4 \cdot 10^6$ \\
99.7	& $3.8 \cdot 10^6$ \\
\hline
\hline
\end{tabular}
\end{table}

Integrating over the limited volume of Figure~\ref{fig:vlim_integrand} as per equation~(\ref{eq:vlim}) produces volumetric limits shown in Table~\ref{tbl:vlim}. At 95\% C.L., the presence of a repeating FRB with burst energy distribution given by equations~(\ref{eq:intrinsic_rate}) and (\ref{eq:defaults}) with Poissonian arrival times can be excluded in a volume of $8.4 \cdot 10^6$\,Mpc$^3$. However, the burst arrival time distribution of FRB~121102 is clearly non-Poissonian --- this is investigated in the next section.

\section{Burst time distribution}
\label{sec:weibull}

The method of Section~\ref{sec:method} can be adapted to any time-distribution of bursts, provided that the probability of observing two or more events in the observation period is a readily calculable function of the expected number of events, $\lambda$. This then allows $\lambda_{\lim}$ to be set to the value at which this probability is equal to the desired level of confidence.

The time distribution of bursts from FRB~121102 is clearly non-Poissonian --- this object either has `active' and `inactive' modes, or a `bursty' distribution, over a variety of timescales \citep{2017ApJ...850...76L,2018ApJ...863....2G,2018ApJ...866..149Z}.\footnote{\citet{2018ApJ...866..149Z} note that observational biases may play a role in skewing perception of its activity. For instance, the only article reporting a non-detection at radio wavelengths appears as a Research Note \citep{2018RNAAS...2a..30P}.} \citet{2016MNRAS.458L..89C} provide a general discussion of non-Poissonian statistics in the context of FRB observations.

\citet{2018MNRAS.475.5109O} modelled the distribution of burst arrival times of FRB~121102 as a Weibull distribution. Their fit used data from \citet{2016Natur.531..202S} and \citet{2016ApJ...833..177S}, with single observations lasting of order one hour, and spanning the period from late 2011 to early 2016 (over three years). The fitted index (or shape parameter) $k$ of the Weibull distribution was $k=0.34^{+0.06}_{-0.05}$ with mean rate $R=5.7^{+3.0}_{-2.0}$\,day$^{-1}$. A value of $k=1$ replicates a Poissonian distribution, and $k<1$ implies data which is more clustered, resulting in a greater probability to see both zero and many events. Limiting the presence of a repeating FRB with burst arrival times following a Weibull distribution with $k<1$ therefore requires increasing $\lambda_{\rm lim}$ in comparison to Poissonian values.

\begin{table*}
\caption{Critical expectation values, $\lambda_{\rm lim}$, corresponding to different confidence levels (C.L.) in three cases of burst time distributions: a Poissonian distribution; a Weibull distribution with shape parameter $k$ viewed by a continuous pointing; and a Weibull distribution viewed by a typical time distribution of CRAFT lat50 pointings.}\label{tbl:weibull}
\centering
\begin{tabular}{c | c | c c c | c c c }
\hline
\hline
& \multicolumn{7}{c}{$\lambda_{\rm lim}$} \\
\hline
& Poisson & \multicolumn{3}{c|}{Single continuous pointing} & \multicolumn{3}{c}{CRAFT lat50 pointing} \\
C.L.\ & (k=1) & k=0.29 & k=0.34 & k=0.40 & k=0.29 & k=0.34 & k=0.40 \\
\hline
68\%	& 2.36 & 12.5& 8.08 & 5.7  & 4.6 & 3.8 & 3.2\\
90\%	& 3.89 & 45.0& 25.4 & 15.8  & 12.3 & 8.6 & 6.6 \\
95\%	& 4.84 & 76.5 & 40.8 & 23.9 & 18.4 & 11.9 & 8.7 \\
99.7\%	& 7.83 & 330 & 150 & 75.2 & 57 & 30 & 20 \\
\hline
\hline
\end{tabular}
\end{table*}

The probabilities of viewing a given number of events in the case of a single continuous pointing are derived in \citet{2018MNRAS.475.5109O}. In particular, the probability of viewing zero or one events --- and hence the probability of viewing two or more --- is readily evaluated. The resulting values of $\lambda_{\rm lim}$ required for a given level of confidence are shown in Table~\ref{tbl:weibull} (columns 2--4). For observations spanning multiple pointings, no analytic expression is obtainable, and the exact case must be simulated through Monte Carlo methods. It is also impossible to reduce multiple overlapping pointings to a single effective exposure (see Section~\ref{sec:procedure}), as is the case for a Poissonian distribution. In the limit of many short pointings separated by large intervals however, the expected rate will again become Poissonian.

\subsection{$\lambda_{\rm lim}$ for the CRAFT lat50 survey}

The CRAFT lat50 survey is an intermediate case of observations spanning typically a few hr per pointing, and totalling several tens of days per survey field, over a one-year period. A full calculation of the requisite probabilities of detecting a repeating FRB with burst times following a Weibull distribution therefore would require a dedicated calculation for each and every pointing. To make this tractable, this calculation is performed for a single field only, `G217-50', centred at Galactic coordinates $l=217^{\circ}$, $b=-50^{\circ}$. This field was observed for the greatest duration (45.3 antenna-days after accounting for efficiency losses), with observations spread over 172 periods, averaging $6.3$\,hr each. This field is chosen not only because it has the greatest impact on the volumetric limit, but also because fields with shorter total pointing duration could either more sparsely span the same time period, and thus have a more Poissonian distribution of burst probabilities (and hence lower $\lambda_{\rm lim}$); or be observed with equal regularity, but span a shorter period, and thus tend more toward the single continuous pointing case (and hence higher $\lambda_{\rm lim}$).

In order to simulate the detection probabilities for a Weibull distribution of burst times in terms of $k$ and $R$, care must be taken that $R$ is specified in terms of true days, not days observed. A random series of burst waiting times, and hence burst actual times, is then generated, setting the first event to be much earlier than the first observation. The full series is then generated over the observation period, and bursts occurring during the observation times count as being detected. This process is then repeated $10^4$ times for each combination of $(k,R)$ in order to estimate the probability of detecting any given number of events $n$, allowing the value of $R$, $R_{\rm lim}$, corresponding to a given confidence limit to be obtained. $\lambda_{\rm lim}$ can then be calculated by multiplying $R_{\rm lim}$ by the total observation time $T_{\rm obs}$. These values are shown in Table~\ref{tbl:weibull}. While the effects of clustering still act to require a higher number of expected events before the presence of an FRB can be excluded at a given confidence level, the required values are much lower than in the case of a single continuous pointing.

\begin{figure}
\begin{center}
\includegraphics[width=\columnwidth]{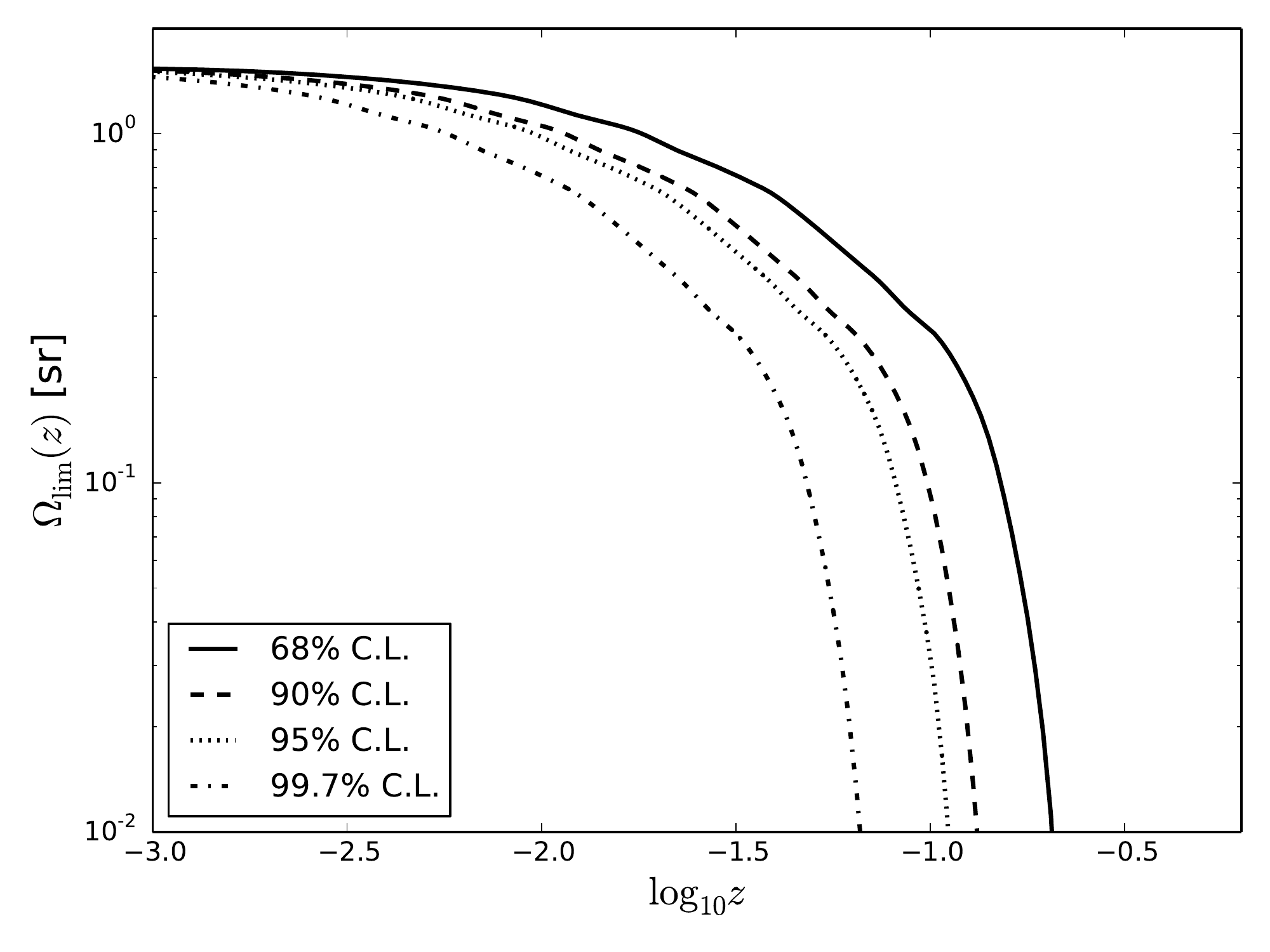}
\caption{
Solid angle $\Omega_{\rm lim}(z)$ over which the presence of a repeating FRB with properties given by equations~(\ref{eq:intrinsic_rate}) and (\ref{eq:defaults}) can be excluded as a function of $z$ at the stated confidence levels (C.L.), assuming a Weibull distribution of burst arrival times with shape parameter $k=0.34$. Calculations use truncated beams with best-case sidelobes.
} \label{fig:zlim_hist_w}
\end{center}
\end{figure}

Using these modified values of $\lambda_{\rm lim}$ for $k=0.34$, and repeating the calculations of Section~\ref{sec:craft}, produces limiting solid angles $\Omega_{\rm lim}(z)$ given in Figure~\ref{fig:zlim_hist_w}. As expected, $\Omega_{\rm lim}(z)$ is contracted to smaller values of $z$ when compared to the Poissonian case (Figure~\ref{fig:zlim_hist}). The resulting values of $V_{\rm lim}$, after applying equation~(\ref{eq:vlim}), are given in Table~\ref{tbl:varying} (second column).

\section{Varying properties of the repeater}
\label{sec:varying}

The properties of FRB~121102 are hardly well-constrained, and furthermore, it would be astounding if all repeating FRBs in the Universe repeated with identical properties to that of FRB~121102. The simplest variation is to consider other values of the standard parameters of equation~(\ref{eq:defaults}). Given that $R_0$ and $E_0$ are degenerate, $E_0$ is kept constant, and the analysis of Section~\ref{sec:weibull} is repeated for different values of $R_0$ and $\gamma$, and the shape parameter of the Weibull distribution, $k$. In each case, other values of the parameters are fixed to their standard values, i.e.\ $R_0=7.4\,{\rm day}^{-1}$, $\gamma=-0.9$, and $k=0.34$.

\begin{figure}
\begin{center}
\includegraphics[width=\columnwidth]{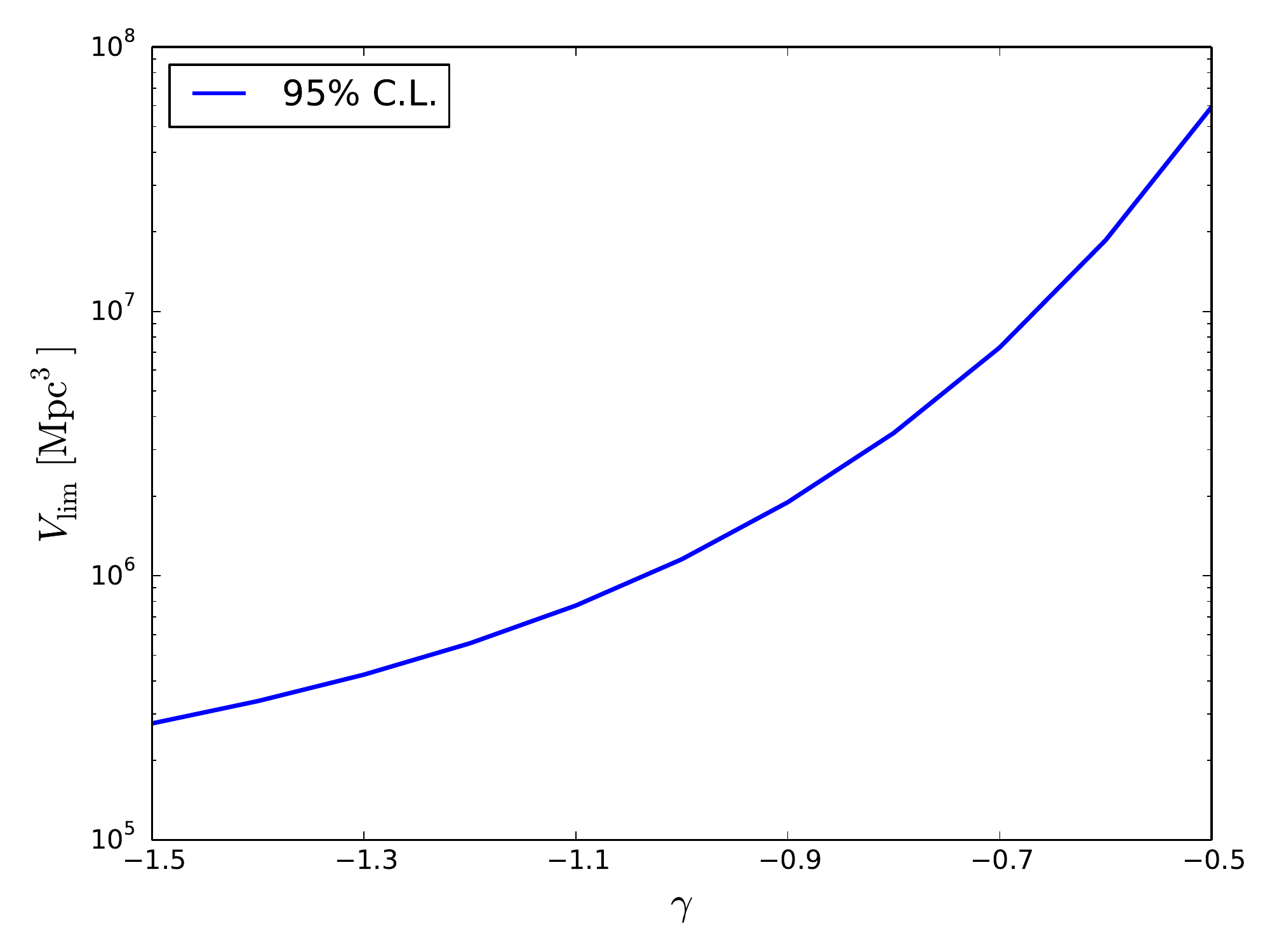}
\caption{
Dependence of $V_{\rm lim}$ on $\gamma$ from equation~(\ref{eq:intrinsic_rate}), with other parameters at default values from equation~(\ref{eq:defaults}).
} \label{fig:gamma}
\end{center}
\end{figure}

\begin{figure}
\begin{center}
\includegraphics[width=\columnwidth]{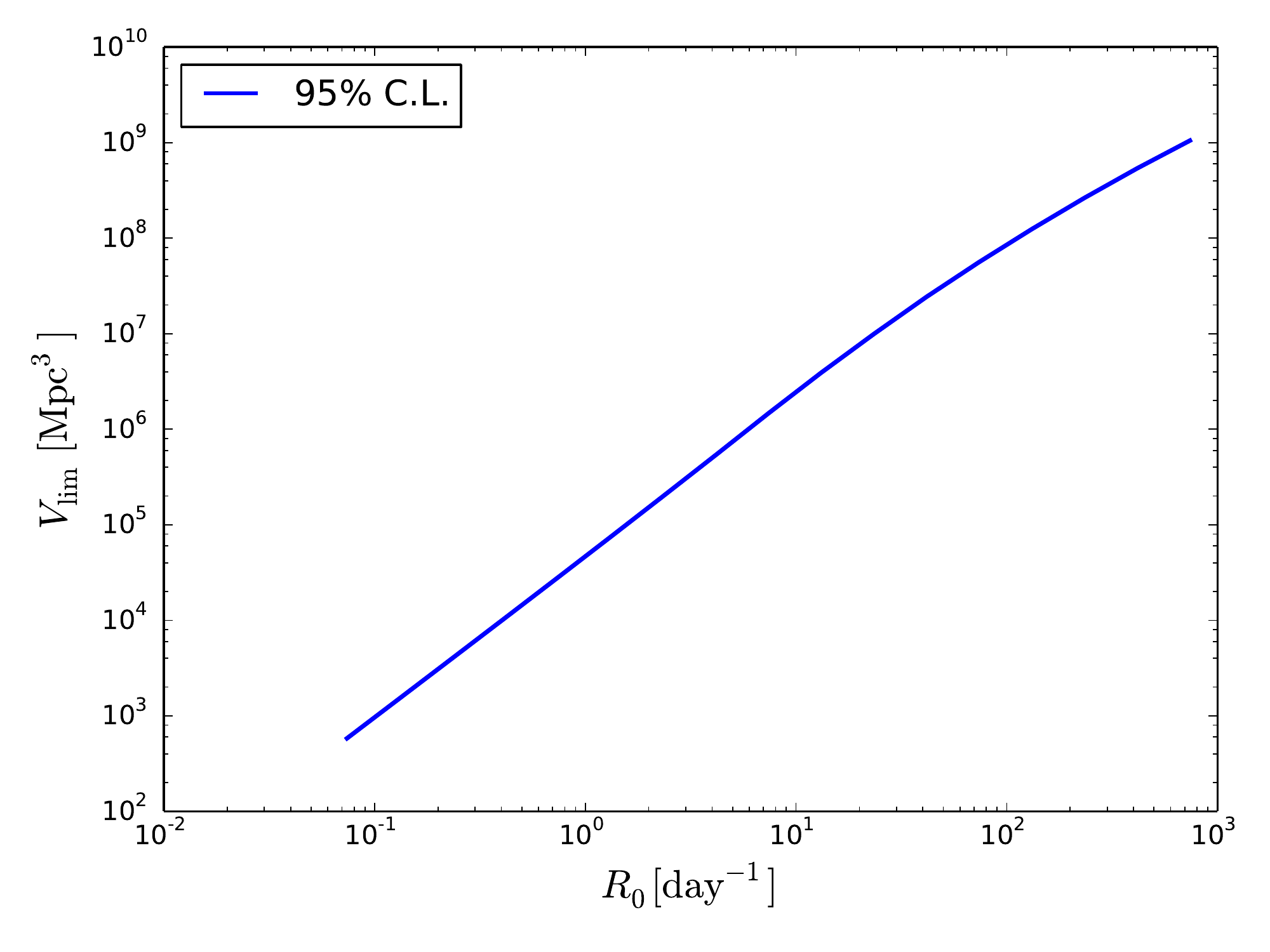}
\caption{
Dependence of $V_{\rm lim}$ on $R_0$ from equation~(\ref{eq:intrinsic_rate}), with other parameters at default values from equation~(\ref{eq:defaults}).
} \label{fig:R0}
\end{center}
\end{figure}

\begin{figure}
\begin{center}
\includegraphics[width=\columnwidth]{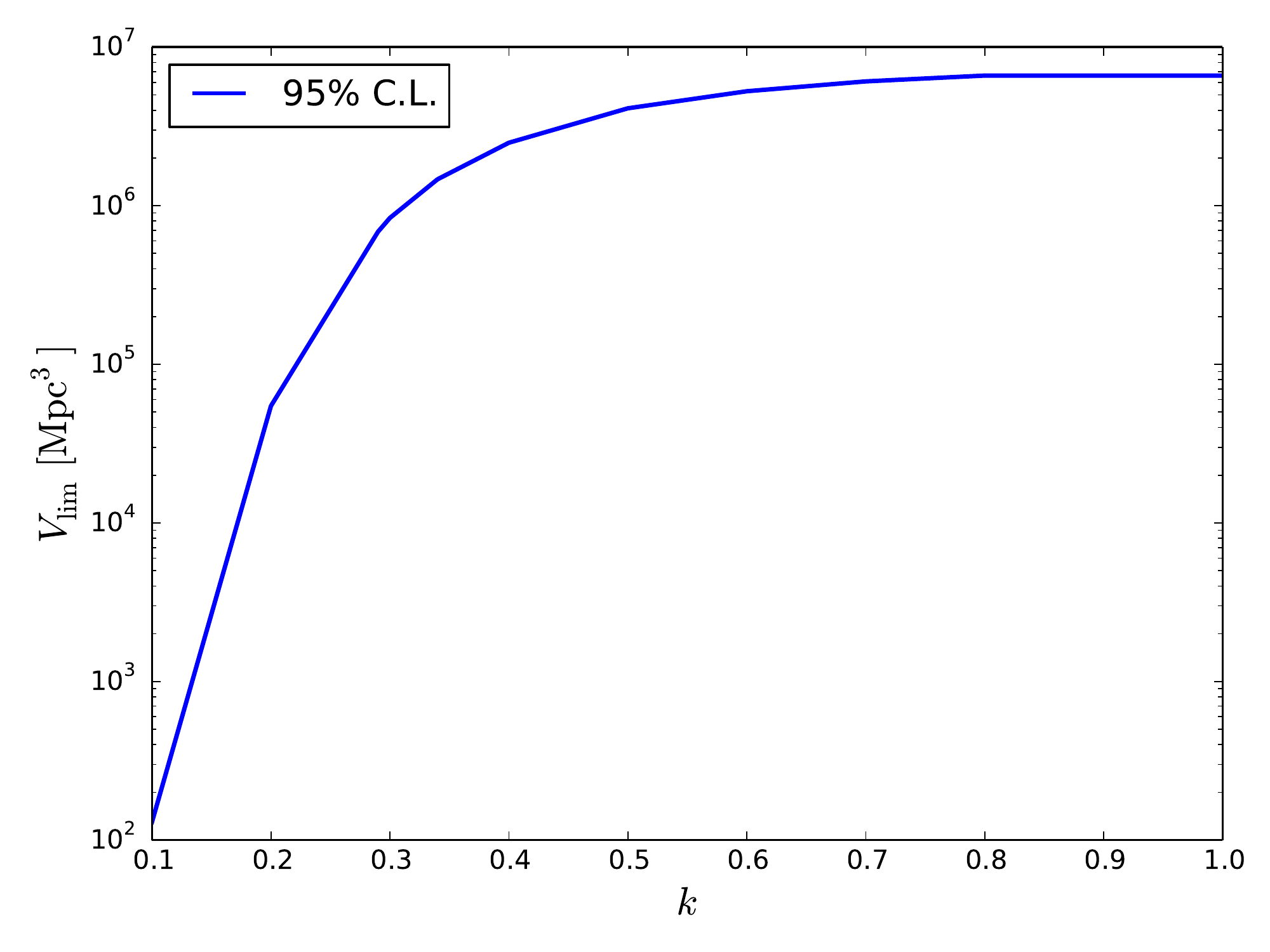}
\caption{
Dependence of $V_{\rm lim}$ on the Weibull shape parameter $k$ from equation~(\ref{eq:intrinsic_rate}), holding other parameters constant. Note that $k=1.0$ corresponds to a Poisson distribution of burst arrival times.
} \label{fig:k}
\end{center}
\end{figure}

Figure~\ref{fig:gamma} shows the 95\% C.L.\ upper limits on $V_{\rm lim}$ resulting from varying $\gamma$ in the range $-0.5$ to $-1.5$; Figure~\ref{fig:R0} does the same when varying $R_0$ (the rate above $1.7 \cdot 10^{38}$\,erg) between $0.074$ and $740$ per day; while Figure~\ref{fig:k} varies $k$ between $0.1$ and $1.0$ (i.e.\ a Poisson distribution). Specific values of $V_{\rm lim}$ for the ranges of $R_0$, $\gamma$, and $k$ consistent with FRB~121102 are given in Table~\ref{tbl:varying}.

Since $\gamma$ determines the trade-off between sensitivity and observing time (i.e.\ how much longer one has to wait to view very bright events), steeper (more negative) values of $\gamma$ reduce the ability of the long ASKAP/CRAFT pointings to probe large volumes of the distant universe using rare, very bright pulses. However, they do not greatly reduce the ability of short observations to observe regular weak pulses from repeating FRBs in the local universe. The variation in $V_{\rm lim}$ for $-1.5 < \gamma < -0.5$ nonetheless covers two orders of magnitude.

The dependence of limiting volume on $R_0$ is very strong. Since $\gamma$ is almost unity, reducing the intrinsic rate by an order of magnitude requires a reduction in observable intrinsic burst energy of almost the same factor through equation~(\ref{eq:intrinsic_rate}). This then reduces $z_{\rm lim}$ almost with the square root of $R_0$, and hence $V_{\rm lim}$ varies approximately as $R_0^{1.5}$, which is clearly observed in Figure~\ref{fig:R0}. The reduction in slope for large $R$ is due to the universe at $z>1$ being probed, where cosmological effects become important.

Finally, as $k$ is reduced from the Poissonian value of $1$, burst arrival times become more clustered. This increases both the probability of seeing zero and many events at the expense of viewing a few. For values of $0.8 \le k \le 1$, the timing of the observations results in estimated values of $\lambda_{\rm lim}$, and hence $V_{\rm lim}$, being identical to within simulation accuracy of the Poissonian case ($k=1$). However, as $k$ becomes small, very large values of $\lambda_{\rm lim}$ are required to exclude the possibility of viewing no repeat events. For example, for $k=0.1$ at $95\%$ C.L., $\lambda_{\rm lim} \sim 2900$, and only a tiny local volume of $100$\,Mpc$^3$ is probed. Since the probability of viewing a single event becomes negligible in the case of low values of $k$, this suggests that the best limits on very bursty FRBs will be derived from the observation of single bursts, rather than the non-observation of repeating bursts as performed here.

\begin{table*}
\caption{
Values of $V_{\rm lim}$ [Mpc$^3$] for repeating FRBs over the range of parameter values compatible with FRB~121102. The units of $R_0$ are days$^{-1}$.
}\label{tbl:varying}
\centering
\begin{tabular}{c | c | c c | c c | c c }
\hline
\hline
C.L.\ & $R_0=7.4$, $\gamma=-0.9$, $k=0.34$ & $R_0=2.6$	&  $R_0=11.4$	& $\gamma=-0.7$	& $\gamma=-1.1$	& $k=0.29$ & $k=0.40$ \\
\hline
68\%	& $1.2 \cdot 10^7$ & $1.7 \cdot 10^6$ & $1.9 \cdot 10^7$ & $4.8 \cdot 10^7$ & $3.2 \cdot 10^6$ &  $7.2 \cdot 10^7$ &  $1.3 \cdot 10^7$ \\
90\%	& $3.3 \cdot 10^6$ & $4.2 \cdot 10^5$ & $5.2 \cdot 10^6$ & $1.0 \cdot 10^7$ & $9.9 \cdot 10^5$ &  $1.4 \cdot 10^6$ &  $4.3 \cdot 10^6$ \\
95\%	& $1.9 \cdot 10^6$ & $2.4 \cdot 10^5$ & $3.1 \cdot 10^6$ & $5.4 \cdot 10^6$ & $6.2 \cdot 10^5$ &  $6.9 \cdot 10^5$ &  $2.6 \cdot 10^6$ \\
99.7\%	& $3.8 \cdot 10^5$ & $4.9 \cdot 10^4$ & $6.2 \cdot 10^5$ & $7.4 \cdot 10^5$ & $1.7 \cdot 10^5$ & $9.8 \cdot 10^4$ &  $6.3 \cdot 10^5$ \\
\hline
\hline
\end{tabular}
\end{table*}

\section{Limits on the population density of repeating FRBs}
\label{sec:volume}

The focus of the previous sections has been on ruling out the presence of a single repeating FRB in a given volume $V_{\rm lim}$. The simplest resulting limit on the FRB population density $\Phi$ (FRBs\,Gpc$^{-3}$), $\Phi_{\rm lim}$, is given by:
\begin{eqnarray}
\Phi_{\rm lim} & = & \frac{\Lambda_{\rm lim}}{V_{\rm lim}}. \label{eq:simple_rholim}
\end{eqnarray}
Here, $\Lambda_{\rm lim}$ is an expectation value for the number of repeating FRBs in $V_{\rm lim}$, and is analogous to $\lambda_{\rm lim}$ defined in Section~\ref{sec:pmult} for the number of bursts from a given FRB. It should be chosen to match the desired confidence level. In this case however, it is relatively safe to assume that the actual number of FRBs inside $V_{\rm lim}$ follows a Poissonian distribution, so that e.g.\ $\Lambda_{\rm lim}=4.84$ corresponds to a 95\% C.L.\ upper limit.

The simple method above ignores that limits will be more stringent when considering larger volumes over which the detection probability of individual repeating FRBs is smaller. Furthermore, it cannot account for changing population densities with redshift, i.e.\ $\Phi \to \Phi(z)$.

The probability $p_{\ge 2}(F_{\rm th},z)$ of detecting two or more bursts from a given FRB depends on both the distance $z$ and the beam sensitivity $B$, which governs the detection threshold $F_{\rm th}=F_0/B$ as discussed in Section~\ref{sec:solid_angle}. The expected number of observed repeating FRBs, $\Lambda_{\rm rep}$, in a given survey is then given by:
\begin{eqnarray}
\Lambda_{\rm rep} & = & \int_0^{\infty} dz \Phi(z) \epsilon(z) D_H \frac{(1+z)^2 D_A^2}{E(z)} \nonumber \\
\epsilon(z) & = & \int_0^{1} dB  p_{\ge 2}(B,z)  \Omega(B), \label{eq:lambda_obs}
\end{eqnarray}
where $\epsilon(z)$ weights the sensitive solid angle over beam sensitivity according to the detection probability. Unlike equation~(\ref{eq:vlim}), the integral in equation~(\ref{eq:lambda_obs}) extends over all $z$, allowing for many repeating FRBs at large distances to contribute to the expected number of events. The factor $p_{\ge 2}(B,z)$ in the integrand gives different weights per unit volume --- generally, this will be close to unity for $z \lesssim z_{\rm lim}$, and fall more rapidly than increasing volume beyond this.

To calculate $p_{\ge 2}(B,z)$, it is most useful to first calculate the expected number of bursts, $\lambda(B,z)$, so that $p_{\ge 2}(B,z) \equiv p_{\ge 2}(\lambda(B,z))$. $\lambda(B,z)$ can be found for the burst energy distribution of equation~(\ref{eq:intrinsic_rate}) by inverting equation~(\ref{eq:zlim}), with $z_{\rm lim} \to z$ and $\lambda_{\rm lim} \to \lambda$, i.e.:
\begin{eqnarray}
\lambda(B,z) = & T_{\rm obs} R_0 \left[ \frac{4 \pi F_0 \Delta \nu_{\rm FRB}}{B E_0} \frac{ D_L(z_{\rm lim})^2}{(1+z_{\rm lim})^{2+\frac{1}{\gamma}}} \right]^{\gamma} ~[z \le z_{\rm crit}] \nonumber \\
& T_{\rm obs} R_0 \left[ \frac{4 \pi F_{0} \Delta \nu_{\rm obs}}{B E_0} \frac{ D_L(z_{\rm lim})^2}{(1+z_{\rm lim})^{1+\frac{1}{\gamma}}} \right]^{\gamma} ~[z > z_{\rm crit}]. \label{eq:lambda_bz}
\end{eqnarray}
In the case of a Poissonian distribution, $p_{\ge 2}(\lambda)$ is given simply by:
\begin{eqnarray}
p_{\ge 2}(\lambda) & = & 1-(1 + \lambda) e^{-\lambda},
\end{eqnarray}
although the result will be different for different distributions (see e.g.\ Section~\ref{sec:weibull}).

Evaluating $p_{\ge 2}$ allows equation~(\ref{eq:lambda_obs}) to be evaluated for any hypothesised $\Phi$. It is useful to write $\Phi$ as the product of a $z$-dependent factor $\phi(z)$ and its value in the current epoch, $\Phi_0$, i.e.:
\begin{eqnarray}
\Phi(z) & = & \Phi_0 \phi(z) \label{eq:rhoz} \\
\Phi_0 & \equiv & \Phi(z=0) \nonumber \\
\phi(z)& \equiv & \frac{\Phi (z)}{\Phi_0}. \nonumber
\end{eqnarray}
This allows limits on $\Phi_0$ to be set for any given observed number of repeating FRBs by choosing appropriate $\Lambda_{\rm rep}=\Lambda_{\rm lim}$. The limit $\Phi_{\rm lim}$ on $\Phi_0$ can be calculated as:
\begin{eqnarray}
\Phi_{\rm lim} & = & \frac{\Lambda_{\rm lim}}{\int_0^{\infty} dz \phi(z) \epsilon(z) D_H \frac{(1+z)^2 D_A^2}{E(z)}}. \label{eq:rho_lim}
\end{eqnarray}
The implicit assumption here is that $\Phi_{\rm lim}$ will be an upper limit on the FRB density. However, this need not be the case, and both upper and lower limits could be calculated by choosing appropriate $\Lambda_{\rm lim}$ should one or more repeating FRBs be detected, as is the case for the recent results from the Canadian Hydrogen Intensity Mapping Experiment (CHIME; \citet{2019Natur.566..235C}).

\subsection{Volumetric limits from the ASKAP/CRAFT lat50 survey}
\label{sec:lat50_volume}

\begin{figure}
\begin{center}
\includegraphics[width=\columnwidth]{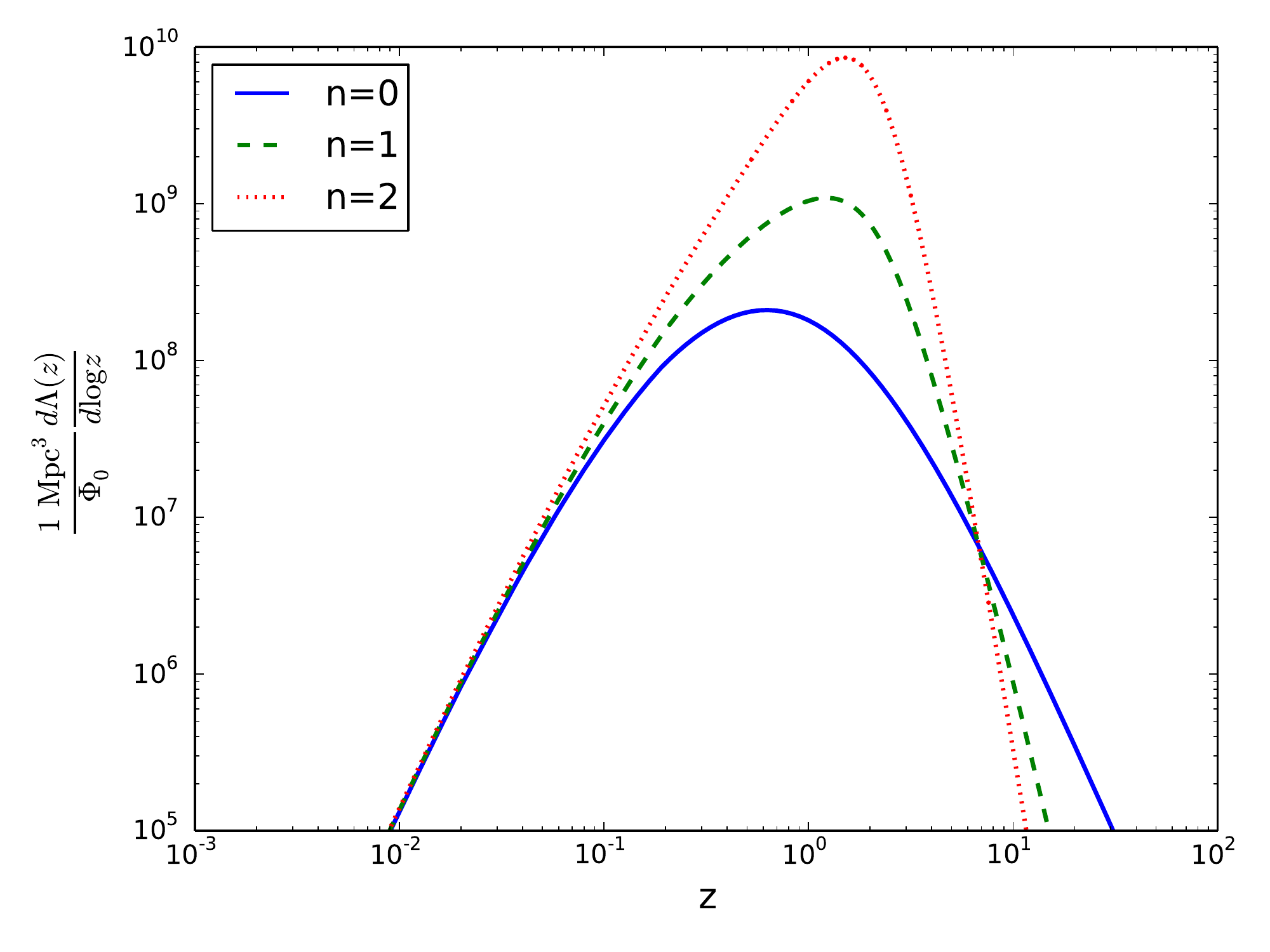}
\caption{
Expected rate of detection $\lambda(z)$ of repeating FRBs similar to FRB~121102 as a function of redshift $z$, for the standard parameter set $R_0=7.4$\,day$^{-1}$, $\gamma=-0.9$, $k=0.34$. The normalised FRB population density $\phi(z)$ is assumed constant ($n=0$; solid), or scales as the star-formation rate to the power of $n=1$ (dotted) and $n=2$ (flat).
} \label{fig:lambda_z}
\end{center}
\end{figure}

In order to limit $\Phi_0$, the z-dependence of the FRB population density $\phi(z)$ is chosen to be an integer power $n$ of the star-formation rate. As per \citet{2018MNRAS.474.1900M}, values of $n=0$ (no dependence), $n=1$ (stellar origin), and $n=2$ (approximately tracing AGN activity) are considered. Evidently this does not encompass models with significant delay-time, e.g.\ long-duration mergers, as tested by \citet{2018ApJ...858...89C}.

The normalised star formation rate (SFR) is taken from \citet{2014ARA&A..52..415M}, giving:
\begin{eqnarray}
\phi(z)& = & \left[ 1.0026 \frac{(1+z)^{2.7}}{1 + \left(\frac{1+z}{2.9}\right)^{5.6}} \right]^n.
\end{eqnarray}
For the standard parameter set, $n=0,1,2$, and using $\Phi_0=1$\,Mpc$^{-3}$, the integrand in equation~(\ref{eq:lambda_bz}) is shown in Figure~\ref{fig:lambda_z}. This is proportional to the repeating FRB detection rate when $\Phi_0=1$\,Mpc$^{-3}$. Since overall dependence on $k$ is not strong (as will be shown below), calculations were made tractable by ignoring the specific time-distribution of ASKAP/CRAFT observations, and calculating $p_{\ge 2}$ assuming that observations on all fields consisted of a single continuous block.

For $n=0$, the distribution peaks near $z=0.6$, rising to $z=1.2$ for $n=1$ and $z=1.5$ for $n=2$. Furthermore, while using a clustered burst time distribution ($k<1$) increases the chance of seeing no events when the expected rate is high (i.e.\ in the near universe), it also increases the chance of seeing two or more events when the expected rate is low (i.e.\ in the distant universe). The first effect reduces the volume in which the presence of a single FRB can be excluded with high confidence, as shown in Section~\ref{sec:weibull}. However, when considering the total detection rate of repeating FRBs, the second effect actually \emph{increases} the expected number. This is because the dominant contribution arises from the larger population at high redshift, where the expected number of bursts $\lambda$ for a given FRB is small, and the probability of observing two or more bursts increases as $k$ decreases.

However, the high redshifts probed in Figure~\ref{fig:lambda_z} illustrate a deficit in the chosen burst energy distribution given in equation~(\ref{eq:intrinsic_rate}): there is no maximum burst energy. \citet{2018Natur.562..386S} finds maximum FRB energies near $10^{33}$\,erg\,Hz$^{-1}$ ($10^{42}$\,erg assuming a 1\,GHz emission bandwidth), while \citet{2018MNRAS.481.2320L} find maximum burst energies of $2 \cdot 10^{44}$\,erg\,s$^{-1}$ ($2 \cdot 10^{41}$\,erg assuming a 1\,ms duration). A two-population scenario where repeating FRBs are only responsible for the weaker observed bursts will result in even lower energy cut-offs.

The nature of any cut-off is not well-constrained, so a simple, sharp cut-off at energy $E_{\rm cut}$ is used, modifying the expected event rate of equation~(\ref{eq:intrinsic_rate}) to be:
\begin{eqnarray}
R(E>E_0) & = &  R_0 \left[ \left( \frac{E}{E_0} \right)^{\gamma} - \left( \frac{E_{\rm cut}}{E_0} \right)^{\gamma} \right]. \label{eq:cut_rate}
\end{eqnarray}
This then modifies equation~(\ref{eq:lambda_bz}) by subtracting $(E_{\rm cut}/E_0)^{\gamma}$ from the $[\ldots]^\gamma$ term. The inclusion of a cut-off renders numerical evaluation significantly more complex --- without such a cut-off, factors of $T_{\rm obs} R_0 F_{\rm th}^{\gamma}$ for any given survey can be pre-calculated in equation~(\ref{eq:lambda_bz}) for all $z$, reducing the problem by one dimension.

\begin{figure}
\begin{center}
\includegraphics[width=\columnwidth]{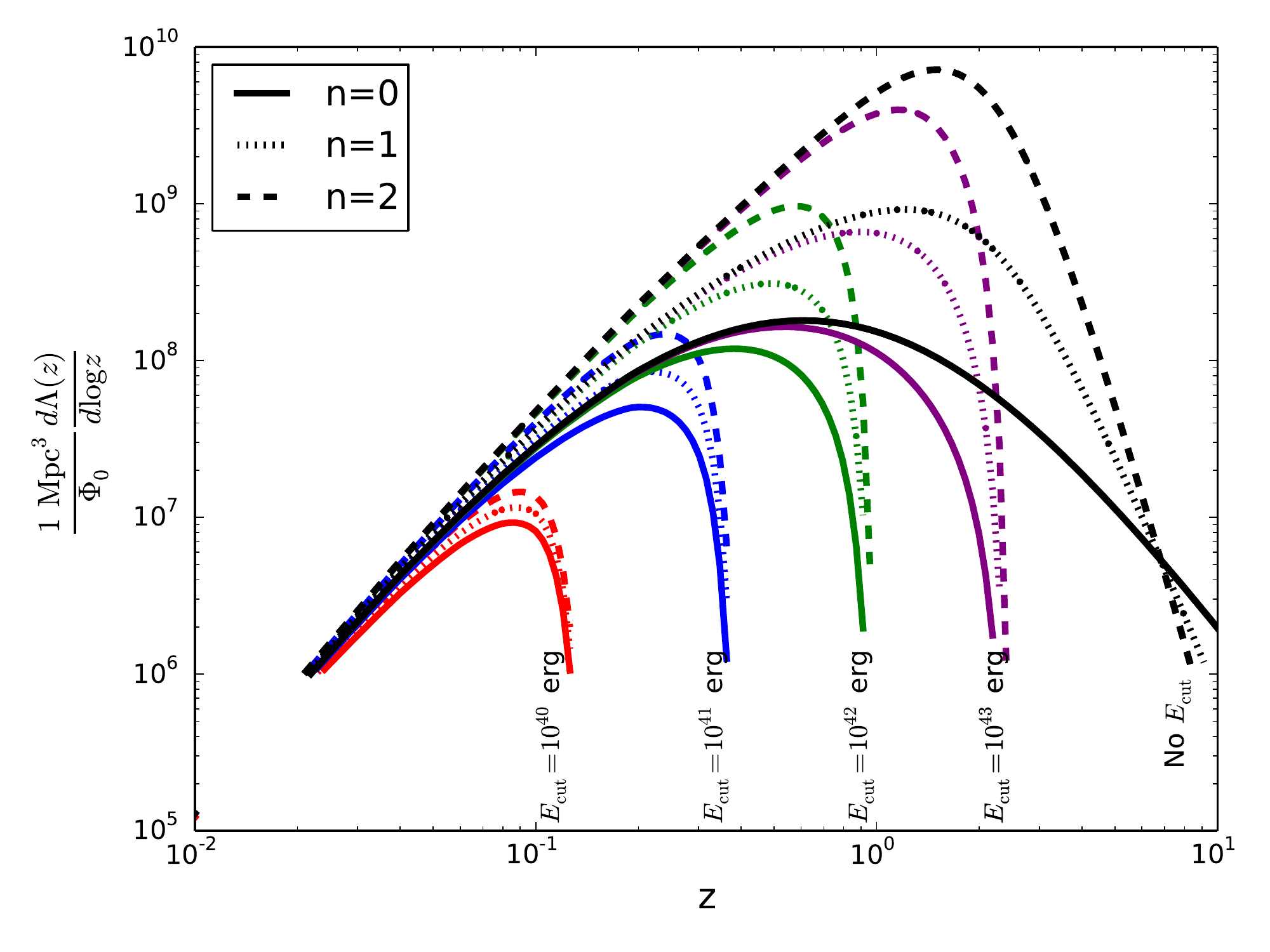}
\caption{
Differential rate of detection of repeating FRBs similar to FRB~121102 as a function of redshift $z$, for the standard parameter set $R_0=7.4$\,day$^{-1}$, $\gamma=-0.9$, $k=0.34$, normalised by $\Phi_0$.  Different colours assume different maximum FRB burst energies $E_{\rm cut}$ (see labels). The normalised FRB population density $\phi(z)$ is assumed constant ($n=0$; solid), or scales as the star-formation rate to the power of $n=1$ (dotted) and $n=2$ (flat).
} \label{fig:Ecut_lambda_z}
\end{center}
\end{figure}

\begin{table}
\caption{
Upper limits at 95\% C.L.\ on $\Phi_0$ (repeating FRBs per Gpc$^3$ in the current epoch) for different assumptions about the maximum burst energy $E_{\rm cut}$ and population evolution (scaling with the star formation rate to the power $n$).
}\label{tbl:rholim}
\centering
\begin{tabular}{c | c | c c | c c | c c }
\hline
\hline
$E_{\rm cut}$ [erg] & $n=0$  & $n=1$ & $n=2$ \\
\hline
$10^{40}$ & 540	& 450 	& 380	\\
$10^{41}$ & 82	& 54	& 34	\\
$10^{42}$ & 27	& 12	& 4.8	\\
$10^{43}$ & 15	& 4.7	& 1.1	\\
None &	12 & 3.0 & 0.55 \\
\hline
\hline
\end{tabular}
\end{table}

Recalculating the limits produces Figure~\ref{fig:Ecut_lambda_z}, with corresponding 95\% upper limits on the population density (i.e.\ $\Lambda_{\rm rep}=4.84$ in equation~(\ref{eq:lambda_bz})) given in Table~\ref{tbl:rholim}. Values of $E_{\rm cut}$ between $10^{40}$\,erg (where pulses from FRB~121102 have been observed \citep{2017ApJ...850...76L}) and $10^{43}$\,erg (well above all known pulse strengths, and beyond which values take their infinite limits) were considered. The result at $10^{42}$\,erg is taken to be the `nominal' value, i.e.\ consistent with all FRBs being due to repeaters similar to FRB~121102.

As expected, the effect of a strong dependence on star-formation rate increases for high values of $E_{\rm cut}$, since this allows the more-distant universe to be probed. For $E_{\rm cut}=10^{42}$\,erg --- close to the derived value for all FRBs --- limits vary by a factor of $2$ for each successive $n$, suggesting that ASKAP observations of repeating FRBs could limit the population evolution model.

\begin{table*}
\caption{
Upper limits at 95\% C.L.\ on $\Phi_0$ (repeating FRBs per Gpc$^3$ in the current epoch) for different assumptions about FRB parameters $R_0$, $\gamma$, and $k$, as a function of the population evolution scaling parameter $n$). The energy cutoff $E_{\rm cut}$ is fixed to $10^{42}$\,erg. The standard scenario uses $R_0=7.4$\,day$^{-1}$, $\gamma=-0.9$, $k=0.34$.
}\label{tbl:ecut_varying}
\centering
\begin{tabular}{c | c | c c | c c | c c c }
\hline
\hline
n &  Standard & $R_0=2.6$	&  $R_0=11.4$	& $\gamma=-0.7$	& $\gamma=-1.1$	& $k=0.29$ & $k=0.40$ & $k=1$ \\
\hline
0 & 27	& 68	& 19	& 11	& 70	& 29 	& 26	& 34	\\
1 & 12	& 33	& 8.2	& 4.8	& 35	& 13	& 12	& 19	\\
2 & 4.8	& 14	& 3.2	& 1.9	& 15	& 4.9	& 4.9	& 9.7	\\
\hline
\hline
\end{tabular}
\end{table*}

Fixing $E_{\rm cut}=10^{42}$\,erg, and varying $R_0$, $\gamma$, and $k$ over the uncertainties of FRB~121102, produces limits on the number of repeating FRBs per Gpc$^3$ in Table~\ref{tbl:ecut_varying}. Unlike the case of excluding the presence of an FRB in a certain volume $V_{\rm lim}$, where changes in the same range of parameters vary the resulting excluded volumes over three orders of magnitude (see Table~\ref{tbl:varying}), limits on $\Phi_0$ vary by only a single order of magnitude over the same range of parameters. One key difference is that varying the Weibull shape parameter $k$ only weakly affects these limits --- indeed, limits become \emph{stonger} for $k<1$, although they again become weaker for $k<0.34$. A steeper burst energy distribution ($\gamma=-1.1$) and an intrinsically lower rate ($R_0=2.6$\,day$^{-1}$) result in the weakest limits, at $\Phi_0 < 70$\,FRBs\,Gpc$^{-3}$.

This highlights that while the detection of multiple pulses from any \emph{specific} repeating FRB in a volume of the universe is both far from likely and highly dependent on particular FRB parameters, the chance of detecting multiple pulses from \emph{any} repeating FRB at much farther distances is much higher, and more robust to changes in FRB parameters.

\subsection{Comparison to the singles rate}
\label{sec:expected_single}

This work has so far concentrated on repeat bursts only, the motivation being that single bursts can be explained by a much wider space of FRB parameters. The ASKAP/CRAFT lat50 survey detected 20 single bursts, 19 of which were above the nominal detection threshold of $9.5\,\sigma$ \citep{2018Natur.562..386S}, which begs the question --- can these 19 events arise from the allowed population of repeating FRBs?

Until now, all limits on repeating FRBs have been calculated conservatively, i.e.\ using lowest-sidelobe beamshapes, setting the telescope threshold according to the DM smearing from objects at high z, and ignoring the effects of sidelobes on overlapping survey fields. However, placing the most conservative limit on repeating FRBs may correspond to an upper, lower, or intermediate limit on the expected ratio of single to multiple pulses.

Of the effects noted above, using large (worst-case) sidelobes increases the total sensitivity of the telescope to FRB bursts, but without high sensitivity in any given direction, thus favouring single bursts. Since much of this sensitivity will be in neighbouring survey fields, in reality this will help to discover repeat bursts. However, ignoring this effect --- as per the discussion of Section~\ref{sec:procedure} --- will mean only increased sensitivity to single bursts will be accounted-for.

\begin{figure}
\begin{center}
\includegraphics[width=\columnwidth]{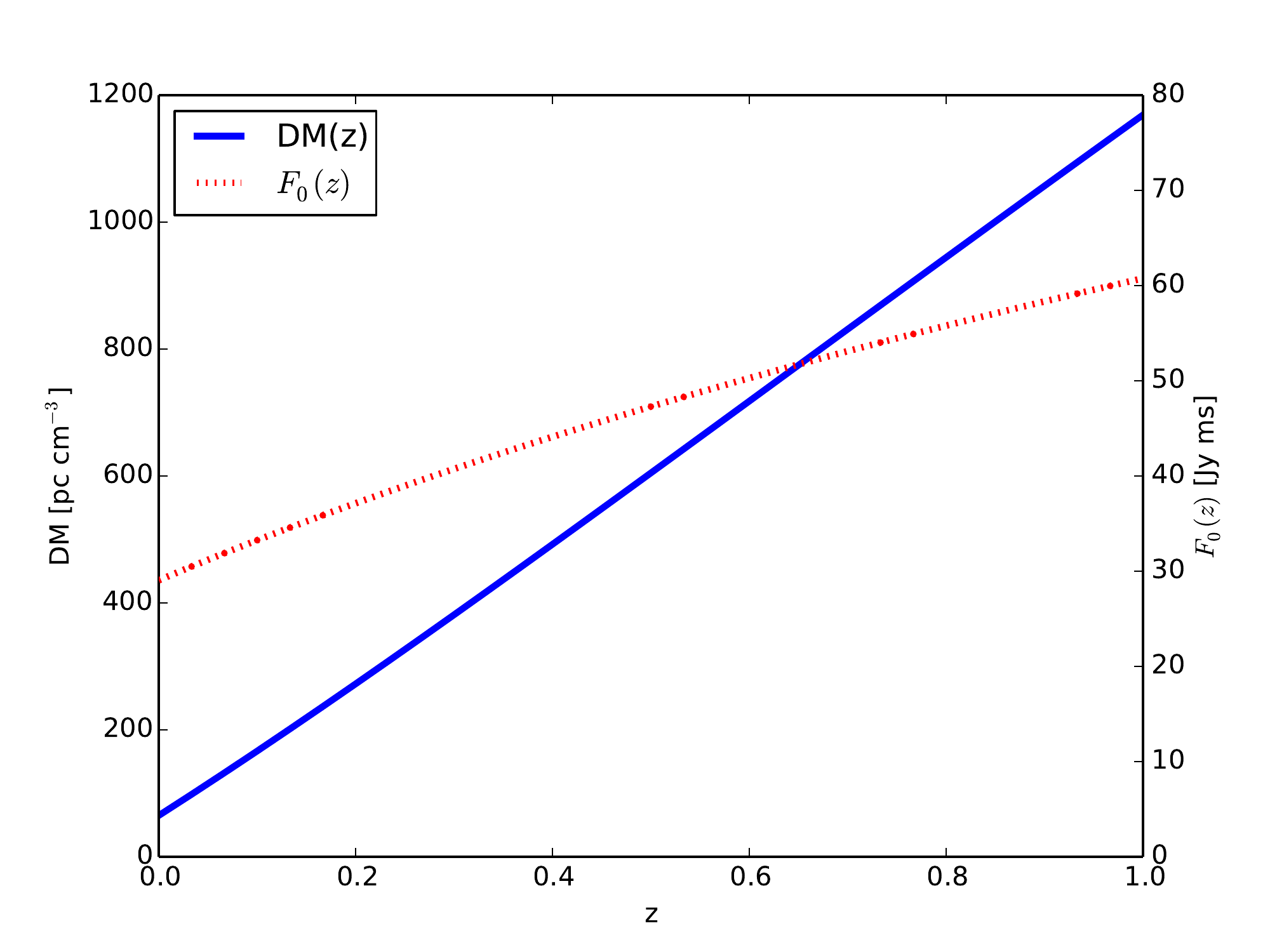}
\caption{
Modelled dependence of dispersion measure DM, and hence detection threshold $F_0$, on redshift $z$.
} \label{fig:dmz}
\end{center}
\end{figure}

Using a redshift-dependent threshold may or may not favour single vs multiple bursts, warranting the extra complexity of modelling it. Therefore, the full $z$-dependent threshold of equation~(\ref{eq:F0DM}) is used, using the treatment of Section~\ref{sec:askap_sensitivity}. The resulting threshold $F_0(z)$ is given in Figure~\ref{fig:dmz}. While this ignores significant contributions from host galaxies and intervening large-scale structures, and breaks down when the IGM ceases to be fully ionised, it should adequately reflect the changing sensitivity to the majority of FRBs in the $z < 1$ range. This $z$-dependent threshold then affects the calculation of the expected number of observed bursts in equation~(\ref{eq:lambda_bz}).

The expected number of single bursts from a repeating FRB population can be calculated by replacing $p_{\ge 2}$ in equation~(\ref{eq:lambda_obs}) with the probability of detecting a single burst, $p_1$, which redefines the efficiency factor $\epsilon(z)$ to $\epsilon_1(z)$:
\begin{eqnarray}
\epsilon_1(z) & = & \int_0^{1} dB  p_{1}(B,z)  \Omega(B). \label{eq:lambda_obs_1}
\end{eqnarray}
Setting $\Phi_{\rm 0}=\Phi_{\rm lim}$ (derived from the non-observation of multiple bursts) then allows the number $\Lambda_1$ of single bursts expected from a maximally allowed population of repeat bursts:
\begin{eqnarray}
\Lambda_{1} & = & \Phi_{\rm lim} \int_0^{\infty} dz \phi(z) \epsilon_1(z) D_H \frac{(1+z)^2 D_A^2}{E(z)}. \label{eq:Lambda1}
\end{eqnarray}
Values of $\Lambda_{1}$ greater than the observed number of single bursts merely indicate that the population of repeating FRBs is significantly less than the 95\% C.L.\ upper limit. However, values of $\Lambda_{1}$ significantly less than that observed require a large population of less repetitive, less bursty repeating FRBs --- including the possibility of a second population of once-off bursts.

Keeping $E_{\rm cut}=10^{42}$\,erg (since any lower value can not by definition account for all singly observed bursts), $\Lambda_1$ is evaluated for the range of repeating FRB parameters investigated in Table~\ref{tbl:ecut_varying}.

Calculations used both best-case and worst-case beamshapes. However, the results differed by less than $1$\% throughout the entire parameter space, and in Table~\ref{tbl:single_bursts} and from hereon, results for best-case beamshapes only are shown.

The effect of a $z$-dependent threshold acted to increase the sensitivity to repeat bursts in comparison to single bursts, i.e.\ the limits in Table~\ref{tbl:single_bursts} are weaker by typically tens of percent when using an (incorrect) constant threshold.

\begin{table*}
\caption{
Maximum expected number of single FRB bursts, $\Lambda_1$, from the ASKAP/CRAFT lat50 survey, assuming a population of repeating FRBs equal to the 95\% C.L.\ upper limits --- and corresponding parameters --- from Table~\ref{tbl:ecut_varying}.
}\label{tbl:single_bursts}
\centering
\begin{tabular}{c | c | c c | c c | c c c }
\hline
\hline
n &  Standard & $R_0=2.6$	&  $R_0=11.4$	& $\gamma=-0.7$	& $\gamma=-1.1$	& $k=0.29$ & $k=0.40$ & $k=1$ \\
\hline
0 & 3.4	& 4.8	& 2.9	& 2.3	& 4.5	& 2.5 	& 4.4	& 14	\\
1 & 4.0	& 5.7	& 3.5	& 2.6	& 5.5	& 3.0	& 5.3	& 18	\\
2 & 4.7	& 6.7	& 4.0	& 3.0	& 6.7	& 3.5	& 6.3	& 24	\\
\hline
\hline
\end{tabular}
\end{table*}

Except for repeating FRBs with Poissonian ($k=1$) arrival times, none of the parameter combinations examined in Table~\ref{tbl:single_bursts} can produce the $19$ single burst events seen by the ASKAP/CRAFT lat50 survey within the limits set in Section~\ref{sec:lat50_volume}. However, the allowed number of single bursts is only a factor of a few below that observed, and could likely be explained by a larger population of repeating FRBs having a lower rate, or less bursty arrival times. While the ratio of single to multiple bursts also increases with increasing (less negative) $\gamma$, extremely flat burst energy distributions ($\gamma \to 0$) seem implausible.

\begin{figure}
\begin{center}
\includegraphics[width=\columnwidth]{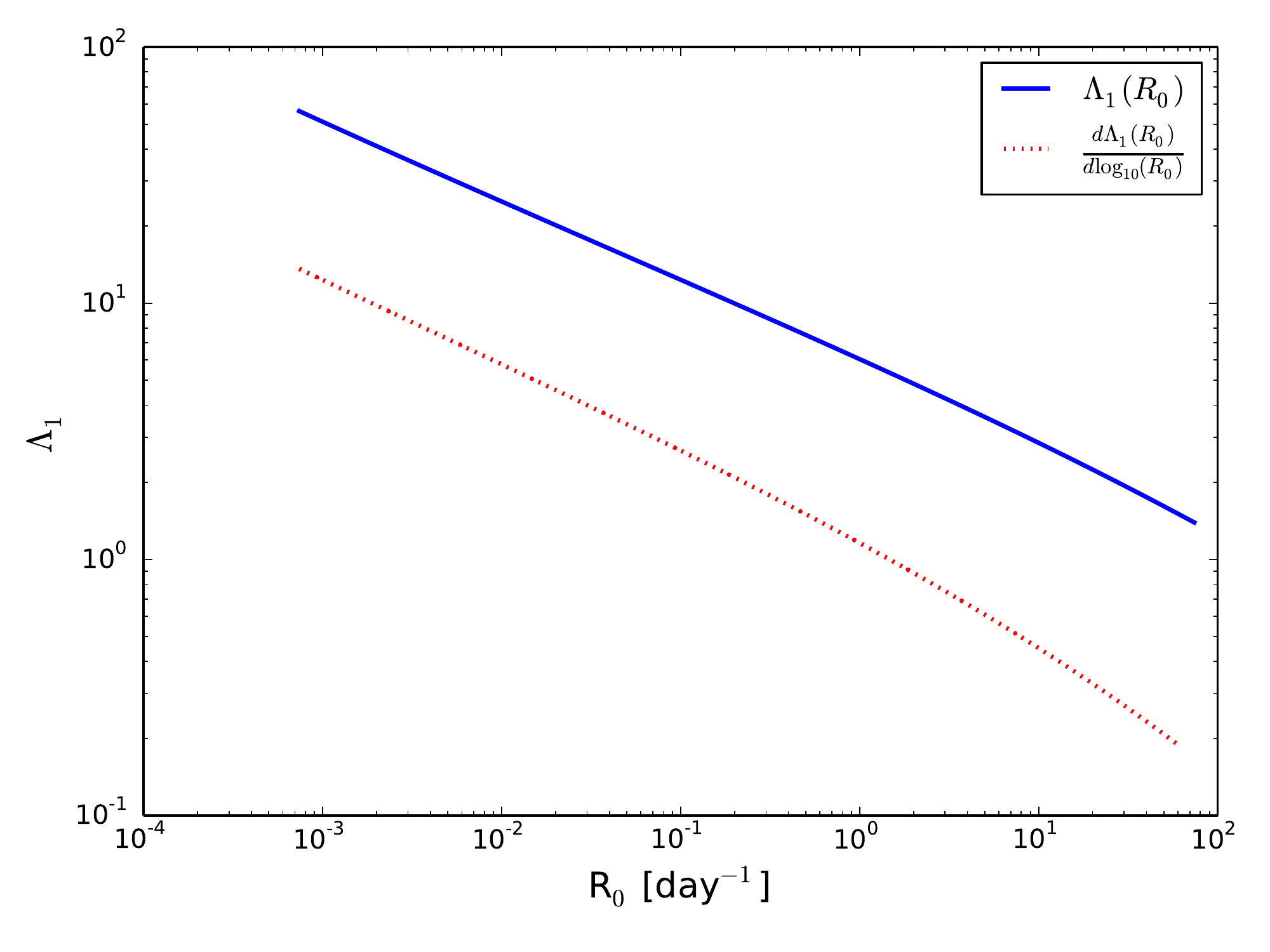}
\caption{
Expected number of single bursts $\Lambda_1$ as a function of repetition rate, $R_0$, for (blue solid line) a population consisting entirely of FRBs with rate $R_0$; and (red dotted line) the differential rate of single bursts from a population consisting of the distribution in $R_0$ from equation~(\ref{eq:diff}). In the latter case, the integral equates to 18 single bursts.
} \label{fig:pop_p1}
\end{center}
\end{figure}

At what rates $R_0$ are the expected number of single bursts $\Lambda_1$ consistent with limits on the population density, $\Phi_{\rm lim}$, and the observed value of 19? In the case of Poissonian arrival times, this is already almost consistent for $R_0=7.4$\,day$^{-1}$. The case of $k=0.34$ is shown in Figure~\ref{fig:pop_p1} --- only a population of repeaters with $R_0 \le 0.02$\,day$^{-1}$ are consistent at 95\% C.L.
This suggests that either the majority of repeaters are less bursty, or repeat less often, than FRB~121102.

\subsection{Comparison to the observed DM/redshift distribution}
\label{sec:dm_comparison}

\begin{figure}
\begin{center}
\includegraphics[width=\columnwidth]{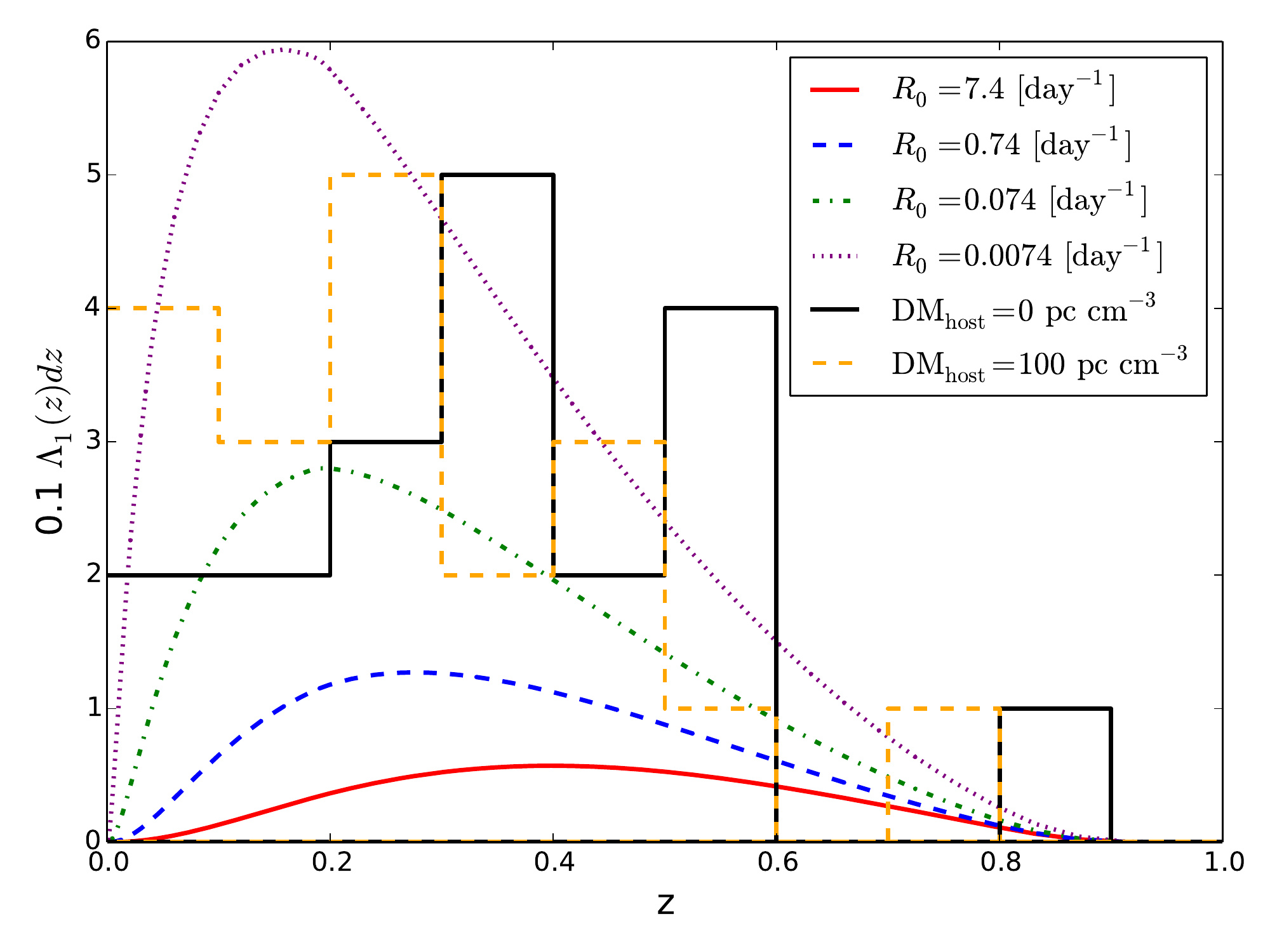}
\caption{
Expected number of single bursts $\Lambda_1$ in the ASKAP/CRAFT lat50 survey as a function of redshift for populations of identical repeating FRBs producing $\Lambda_{\rm rep}=4.84$ repeating sources. Four values of $R_0$ are considered; all other parameters are held constant at $k=0.34$, $\gamma=-0.9$, $E_{\rm cut}=10^{42}$\,erg, and $n=0$. Also plotted are redshift histograms from the ASKAP/CRAFT lat50 survey for 19 above-threshold FRBs, calculated from observed DMs assuming no host DM contribution (black histogram) and a contribution of ${\rm DM}_{\rm host}=0$\,pc\,cm$^{-3}$ (orange histogram).
} \label{fig:pop_z}
\end{center}
\end{figure}

It is also relevant to compare the redshift distribution of observed ASKAP/CRAFT lat50 FRBs with that predicted from the integrand of equation~(\ref{eq:Lambda1}). This is shown in Figure~\ref{fig:pop_z}, for four cases of $R_0$, and standard parameters $k=0.34$, $\gamma=-0.9$, $E_{\rm cut}=10^{42}$\,erg, and $n=0$. The redshifts of ASKAP/CRAFT lat50 FRBs are calculated from the observed DMs \citep{2018Natur.562..386S}, and assuming a total Galactic (including halo) contribution of 65\,pc\,cm$^{-3}$ as per Section~\ref{sec:askap_sensitivity}. Host galaxy contributions, DM$_{\rm host}$, were set at $0$ and $100$\,pc\,cm$^{-3}$ (histograms). The DM$_{\rm host}=0$ case is not consistent with any rate: either the total number of single bursts is under-predicted ($R_0=0.74$ and $7.4$\,day$^{-1}$), or the peak of the redshift distribution is too low ($R_0=0.074$ and $0.0074$\,day$^{-1}$). Assuming DM$_{\rm host}=100$ is generally consistent in both shape and magnitude with the $R_0=0.074$ and $0.0074$\,day$^{-1}$ cases.

\begin{figure}
\begin{center}
\includegraphics[width=\columnwidth]{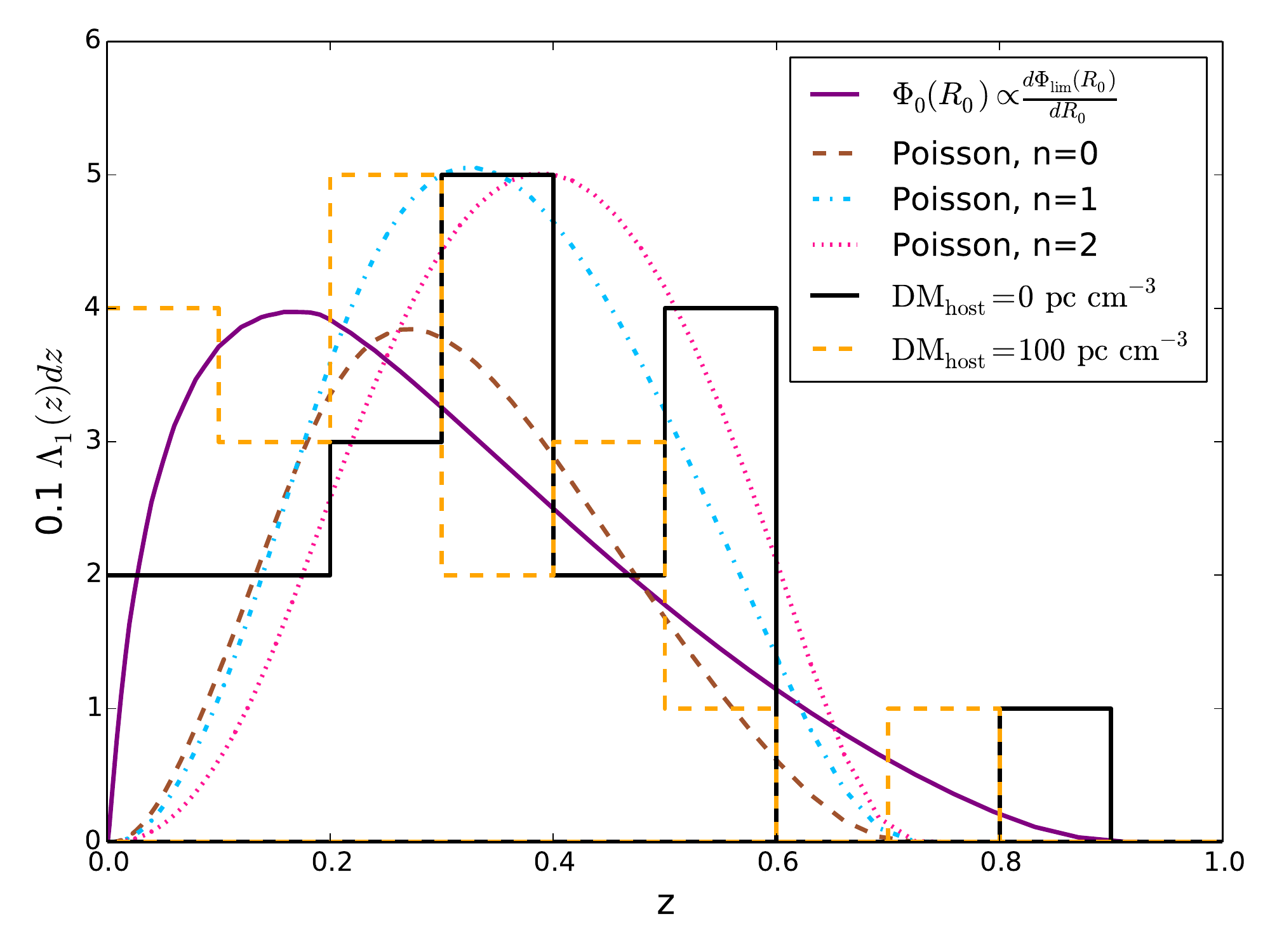}
\caption{
Expected number of single bursts $\Lambda_1$ in the ASKAP/CRAFT lat50 survey as a function of redshift for different repeating FRBs populations. Three cases where all FRBs are identical are considered, with Poissonian burst arrival times ($k=1$), $\gamma=-0.9$, and $E_{\rm cut}=10^{42}$\,erg; only the scaling parameter $n$ of the population density with the star formation rate is varied. A fourth case, being the population of repeating FRBs with $R_0$ distribution described by equation~(\ref{eq:diff_vals}), and $k=0.34$, $\gamma=-0.9$, $E_{\rm cut}=10^{42}$\,erg, and $n=0$ is also shown. In the $n=1$ and $n=2$ cases, the distribution is normalised to $19$ single bursts; in the other two cases, no normalisation is applied. Also plotted are redshift histograms from the ASKAP/CRAFT lat50 survey for 19 above-threshold FRBs, calculated from observed DMs assuming no host DM contribution (black histogram) and a contribution of ${\rm DM}_{\rm host}=0$\,pc\,cm$^{-3}$ (orange histogram).
} \label{fig:pop_z_misc}
\end{center}
\end{figure}

The predictions of Figure~\ref{fig:pop_z} apply only when all FRBs repeat at the same rate. Given the existence of a least one strongly repeating FRB, the FRB population might be better described in terms of a differential population distribution, $\Phi(R,z)=\Phi_0(R_0) \phi(z)$, where $\Phi_0(R_0) d R_0$ describes the density of FRBs at $z=0$ repeating with rate between $R_0$ and $R_0 + dR_0$. Rather than attempting to fit such a population to data, here the shape of $\Phi_0(R_0)$ is chosen to match the strength of limits from equation~(\ref{eq:rho_lim}):
\begin{eqnarray}
\Phi_0(R_0) & \propto & \frac{d \Phi_{\rm lim}(R_0)}{d R_0} \label{eq:diff}
\end{eqnarray}
between (somewhat arbitrary) values of $R_0$ of $74$\,day$^{-1}$ and $0.00074$\,day$^{-1}$. The constant of proportionality is chosen to preserve $\Lambda_{\rm rep}=4.84$ for this population. This produces a population well-fitted by a power-law distribution of rates:
\begin{eqnarray}
\Phi_0(R_0) dR_0 & = & 130\,{\rm Gpc}^{-3} \,\left(\frac{R_0}{1\,{\rm day}^{-1}}\right)^{-2.1} \frac{dR_0}{1\,{\rm day}^{-1}}. \label{eq:diff_vals}
\end{eqnarray}
The single-burst predictions from this model are shown in Figure~\ref{fig:pop_p1}, with a total of $\Lambda_1=18$ single bursts over all $R_0$. The corresponding redshift distribution is shown in Figure~\ref{fig:pop_z_misc}. Both the shape and magnitude agree with lat50 results when DM$_{\rm host}=0$.

An alternative is that the majority of FRBs obey Poisson statistics, also shown in Figure~\ref{fig:pop_z_misc}. The observation of an FRB at $z > 0.8$ is unlikely in this model, ruling out agreement with DM$_{\rm host}=0$, while too few low-z bursts are predicted when DM$_{\rm host}=100$. Changing the dependency on the star-formation rate from $n=0$ to $n=1$ or $n=2$ (also shown) allows more single bursts than observed. Since a population density less than the maximum allowed by the 95\% C.L.\ limit ($\Lambda_{\rm rep}=4.84$) is clearly possible, the predicted number of single bursts have been re-scaled to the observed number of $19$. Doing so only slightly shifts the distribution to the right however, and solves neither problem.

Further modelling of the full FRB population, including more sophisticated models of dispersion measure, and explicit fits to data, is left to a future work.

\section{Discussion}
\label{sec:discussion}

The primary result of this work is that the non-observation of repeating pulses in the ASKAP/CRAFT lat50 survey excludes the presence of any repeating FRBs with properties similar to that of FRB~121102 in a volume of $1.9 \cdot 10^6$\,Mpc$^3$ at 95\% C.L. The volume over which FRBs can be excluded ranges from $2.4 \cdot 10^5$\,Mpc$^3$ when considering FRBs repeating at the lowest rate estimates for FRB~121102 ($R_0=2.6$\,day$^{-1}$), to $5.4 \cdot 10^6$\,Mpc$^3$ when considering FRBs with a flatter distribution of pulse strengths ($\gamma=-0.7$). The limited volume increases to $8.4\cdot10^6$ \,Mpc$^3$ in the case of a Poissonian distribution of burst arrival times. These ranges are not uncertainties in the limited volume --- they are uncertainties in the properties of FRB~121102. Rather, the greatest uncertainties in these limits are due to uncertainties in the ASKAP beamshape, and the treatment of overlapping beamshapes from neighbouring fields. The treatment chosen here is always the most conservative, and a more complete treatment could only act to strengthen the limits. Thus the limits derived here are robust.

The properties of any purported repeating FRB strongly affect the volume over which it can be limited. For instance, extremely bursty FRBs with burst times following a Weibull distribution with shape parameter $k=0.1$ can only be excluded in a volume of $100$\,Mpc$^3$ at 95\% C.L. Conversely, limits on extremely bright FRBs are very strong: no FRBs with a rate of $1000$\,day$^{-1}$ exist within a volume if $1$\,Gpc$^3$ (95\% C.L.). Importantly, limits apply independently at each part of the parameter space, e.g.\ both limits for $k=0.1$ and $R_0=1000$\,day$^{-1}$ apply simultaneously and independently. For practical reasons, limits for only a small range of parameter combinations have been presented here.

Limiting the population of repeating FRBs as a whole becomes sensitive to any high-energy cut-off in the burst rate, and the population evolution with redshift. Of the range of scenarios covered by Table~\ref{tbl:rholim}, there are two `benchmark' cutoffs, $E_{\rm cut}$.  These are $E_{\rm cut}=10^{40}$\,erg, which is consistent with the burst energies observed from FRB~121102, but requires a separate population to explain non-repeating bursts; and $E_{\rm cut}=10^{42}$\,erg, which covers all known FRBs. Respective upper limits at 95\% C.L.\ on the population density $\Phi_0$ at $z=0$ are very low, at $540$\,FRBs\,Gpc$^{-3}$ and $27$\,FRBs\,Gpc$^{-3}$. They are stronger for any evolutionary scenario.

That the number of above-threshold single bursts (19) observed in the ASKAP/CRAFT lat50 survey cannot be explained by repeating populations of FRBs within these limits indicates either a population of less frequently repeating FRBs; FRBs repeating with Poisson statistics; or a second population of once-off bursting objects. Compared with the observed DM ($\sim$redshift) distribution, the only consistent scenarios found were a population of FRBs producing no more than $0.074$ bursts day$^{-1}$ above $1.7 \cdot 10^{38}$\,erg, or a population with a distribution of rates of $d\Phi/dR_0 \propto R_0^{-2.1}$. These both required DM contributions from host galaxies to fit the observed distribution. Populations of repeating FRBs obeying Poisson statistics were less consistent, while invoking an appropriate second population of once-off bursts would also have reproduced observations. What can be ruled out with very high confidence is that FRB~121102 is not a typical FRB: either it repeats when most do not; it repeats much more frequently than is typical; or (less likely) it repeats with much burstier statistics than is typical.

The only comparable study of the population density of repeating FRBs is that by \citet{2019MNRAS.484.5500C}. These authors use burst indices in the range $0 > \gamma > -1$ over the burst energy range $10^{35}$--$10^{43}$\,erg, a broad range of intrinsic rates, a population evolving with the star-formation rate ($n$=1), and both Weibull and Poisson distributions. A Monte Carlo simulation is used, allowing the dispersion measure to be modelled in much greater detail than is performed here. \citet{2019MNRAS.484.5500C} do not model a particular survey; however, a single instance of their simulation for a population of Poissonian-distributed FRBs produces $15$ single bursts and $3$ repeating sources in a hypothetical all-sky survey with Parkes. The true expectation values for single and multiple bursts will thus be in the range $15^{+5}_{-3}$ and $3_{-0.9}^{+2.9}$ respectively (68\% confidence intervals). This is similar to the ratios found here, although the difference in beamshape and total time per pointing make exact comparisons difficult. \citet{2019MNRAS.484.5500C} also claim that the chance of ASKAP observing repeat pulses is ``highly unlikely''. This claim is clearly contradicted, since the non-observation of repeat bursts is shown here to be significant, and constrains the population distribution of repeating FRBs.

\subsection{Notes of caution}

There are three mains caveats to these limits. Firstly, all limits presented in this work apply only in the case of isotropic emission. If, as seems likely, FRBs are beamed with beaming factor $f_b$ (i.e.\ into an angle of $4 \pi/f_b$\,sr), then the limits do not apply to FRBs with emission beamed away from Earth, and they will be a factor of $f_b$ weaker. This assumes a constant beaming direction for all bursts from a given FRB. If the beaming direction is varies burst-to-burst, then the rate $R$ should be interpreted as the observed burst rate, with the true burst rate being higher. In such a case, the nominal rate of $R_0=7.4$\,day$^{-1}$ for FRB~121102 is also the observed rate, and the limits therefore do not change. Such a consideration does \emph{not} apply however to conclusions derived from the ratio of single to multiple bursts.

Secondly, no frequency dependence of the FRB rate or spectrum is considered, although the methods clearly could be adapted to such a consideration. The ASKAP/CRAFT lat50 survey from which these limits derive covered the frequency range $1128$ to $1464$\,MHz, with observed FRBs found to have a burst spectral index, $\alpha$ (fluence $F_{\nu} \propto \nu^\alpha$), of $\alpha=-1.5^{+0.2}_{-0.3}$ \citep{2019ApJ...872L..19M}. Most observations of FRB~121102 (see Appendex~\ref{sec:repeater}) however have been at higher frequencies. If single ASKAP/CRAFT bursts are attributable to repeating FRBs observed only once (e.g.\ the scenarios examined in Section~\ref{sec:volume}), this suggests that the lower frequency of the ASKAP/CRAFT lat50 survey was relatively more sensitive to repeating FRBs, not less. This, and the observation of FRB~180814.J0422+73 at even lower frequencies of $400$--$800$\,MHz \citep{2019Natur.566..230C,2019Natur.566..235C}, suggests that it is unlikely that the frequency range used in this work weakens the derived limits.

It is also known that many bursts --- both from FRB~121102, FRB~180814.J0422+73, and the single bursts detected by ASKAP/CRAFT --- have complex frequency structure. If bursts occupy a narrow frequency range, which varies from burst to burst, then $R_0$ should be interpreted as the rate at which bursts are produced between $1128 (1+z)^{-1}$ and $1464 (1+z)^{-1}$\,MHz, for bursts emitted at redshift $z$. In this case, the fitted burst spectral index of $\alpha=-1.5^{+0.2}_{-0.3}$ reflects not the spectral properties of the bursts themselves, but rather a higher rate of bursts at low frequencies ($R \propto \nu^{\alpha}$). Again, the conclusion is that bursts at low frequencies are \emph{more} likely, and the limits presented here remain strong.

Thirdly, unlike \citet{2018MNRAS.481.2320L} and \citet{2018NatAs...2..839C}, no detailed modelling of DM contributions from FRB host galaxies or intervening structures is performed. The main effect of fully modelling the DM--$z$ relation is to produce a small fraction of FRBs with large excess DM \citep{2014ApJ...780L..33M,2015MNRAS.451.4277D}. By making these FRBs less detectable, the limits presented in Sections~\ref{sec:craft} to \ref{sec:varying} will be slightly weakened to apply only to the  majority of the FRB population. The effect is even smaller when considering the observed number of single bursts in Section~\ref{sec:expected_single}, since FRBs with anonymously high DMs will not contribute to either the single or multiple burst detection probabilities.  Similar considerations apply to scatter broadening, which can also lead to reduced sensitivity for a small subset of the population \citep{2018ApJ...865..147Z}.

Comparisons to the observed dispersion measure distribution in Section~\ref{sec:dm_comparison} must be treated with caution however --- FRBs with anomalously high DMs in ASKAP/CRAFT data will produce incorrectly high redshifts. This is why models predicting too few low-$z$ FRBs compared to data can be better excluded than those failing to predict the single high-DM (nominally $0.8<z<0.9$) burst.

\subsection{Comparison to known repeating FRBs}

The two known repeating fast radio bursts --- FRB~121102, and FRB~180814.J0422+73 --- are located at distances of $z=0.19273$ \citep{2017Natur.541...58C} and $z \lesssim 0.1$ \citep{2019Natur.566..235C}, with enclosed volumes of $2.2$\,Gpc$^3$ and $0.34$\,Gpc$^3$ respectively. The limit on the population density of FRBs at $z=0$ presented in Table~\ref{tbl:rholim} allows for at most $27$\,Gpc$^{-3}$ in the standard scenario. This suggests there may not be very many more nearby objects to be found. That any part of the FRB sky is even fractionally complete seems rather incongruous, given that the history of FRBs is one of a few discoveries per year against rates of thousands of bursts per sky per day. However, the advent of long-duration FRB surveys with wide field of view instruments at high sensitivity has resulted in a significant part of the sky being probed --- almost 1\,sr out to $z=0.03$ in the case of the ASKAP/CRAFT lat50 survey. It would be interesting for limits from other FRB-hunting experiments, in particular those using Parkes and CHIME, to be calculated. All repeating FRBs in the local universe may be discovered in the not-too-distant future.

\subsection{Limits on FRB progenitor models}

Several theories of FRB origin invoke object classes compatible with the long-term repetition properties of FRB~121102 (see e.g.\ \citet{2018NatAs...2..842P,2018arXiv181005836P}), which has been observed to repeat since 2012 \citep{2014ApJ...790..101S}. Of these models, most have no predictions for the expected FRB population density. Two cases however --- models relating to young neutron stars (NS)/magnetars \citep{2010vaoa.conf..129P,2013Sci...341...53T,2014MNRAS.442L...9L,2015ApJ...807..179P,2017ApJ...843L..13Z}, and magnetised white dwarfs (WD) resulting from WD--WD mergers \citep{2013ApJ...776L..39K} --- can be estimated from the approximate population densities of their progenitor systems, allowing meaningful comparisons to limits.

The formation rate of young magnetars can be related to the rate of gamma-ray bursts (GRBs), since both known classes of GRB (hypernovae and compact object mergers) can lead to magnetar formation \citep{2014ApJ...780L..21Z}. Including sub-luminous events, the estimated rate of long-duration GRBs is $2 \cdot 10^3$--$2 \cdot 10^4$\,Gpc$^{-3}$\,yr$^{-1}$ \citep{2007ApJ...657L..73G}, while for short GRBs, it is $1100_{-470}^{+700}$\,Gpc$^{-3}$\,yr$^{-1}$ \citep{2012MNRAS.425.2668C}. The single observed NS--NS merger also corresponds to a rate of ${\mathcal O}\sim10^3$\,Gpc$^{-3}$\,yr$^{-1}$ \citep{2017PhRvL.119p1101A,2018MNRAS.481.1908K}).

If a remnant magnetar (which need not be supramassive) emits repeating pulses for at least ${\mathcal O}\sim10$\,yr (spanning the time period over which FRB~121102 has been observed, and consistent with \citet{2018PASJ...70...39Y}), then the population density of such objects is at least $2 \cdot 10^{4}$\,Gpc$^{-3}$. All combinations of population evolution and energy cut-off studied in Table~\ref{tbl:rholim} exclude this optimistic scenario at 95\% C.L.\ by a factor of at least 40. For $E_{\rm cut}=10^{42}$\,erg and $n=0$, less than one in $600$ such objects could produce FRBs similar to FRB~121102. Estimates of a longer emission time (e.g.\ \citet{2018MNRAS.481.2407M} estimate an age of 30--100\,yr for a magnetar powering FRB~121102) result in a larger population, and hence, greater conflict with these limits.

In other words, either FRB~121102 repeats much more frequently than typical objects of its class, and a large number of less frequently repeating objects abound; or only a very small fraction of GRBs produce magnetars; or a very small fraction of young magnetars produce repeating FRBs; or emission from such FRBs is strongly beamed ($f_b \sim 600$) and FRB~121102 is pointing towards us; or FRBs are not associated with GRBs. The first scenario is consistent with conclusions from the observed dispersion measure distribution of ASKAP/CRAFT FRBs, and by noting that the first discovered object of a population is usually exceptional (i.e.\ repeats more often than average) \citep{2018MNRAS.474.1900M}. Integrating the population distribution of equation~(\ref{eq:diff_vals}) above $R_0=0.074$\,day$^{-1}$ produces $1750$\,FRBs\,Gpc$^{-3}$, consistent with the rate of long-duration GRBs.

In the case of the second scenario, while the fraction of GRBs producing black holes or insufficiently magnetised NS will reduce the expected population density, whether or not these effects can account for a factor of $600$ is yet to be shown. The third scenario raises the new question of why such a small fraction of magnetars would produce repeating FRBs. The last two cases therefore seem the most likely: beaming is already motivated by the exceptional power of the radio pulses themselves, while of course neither hypernovae nor NS--NS mergers may be associated with repeating FRBs.

The total WD--WD merger rate has been argued to be ${\mathcal O}\sim 10^{4}$\,Gpc$^{-3}$\,yr$^{-1}$ by \citet{2013ApJ...776L..39K}, comparable to the observed rate of SN~Ia (e.g.\ \citet{2012PASA...29..447M}). The expected beaming factor of radio emission in the model of \citet{2013ApJ...776L..39K} is approximately $f_b=10$. Again assuming a minimum lifetime of $10$\,yr for subsequent emission (the effect of which is to counter the reduction in event rate due to the beaming factor), all former comments from the magnetar scenario apply, except that the beaming factor is already accounted-for.  It is likely however that only a very small fraction of WD--WD mergers produce massive, magnetised WDs --- if this scenario is correct, that fraction must be as low as 0.1\%.

The extreme rarity of FRBs with bursts as rapid (or equivalently, as strong) as those from FRB~121102 may thus indicate that effects due to the chance alignment of interstellar medium properties, such as plasma lensing \citep{2017ApJ...842...35C} or superradiance \citep{2018MNRAS.475..514H}, are primarily responsible for the emission. While predictions for such occurrences are non-existent, they require the confluence of very specific properties of the intervening medium, with (presumably) a correspondingly low rate.

\section{Conclusions}

A method by which limits can be placed on the number of repeating FRBs in a given volume, and their total population density, has been presented. Applied to the `lat50' survey with ASKAP/CRAFT, in which no repeating pulses were observed, the presence of an FRB with properties similar to FRB~121102 in a volume of $1.9 \cdot 10^6$\,Mpc$^3$ can be excluded at 95\% confidence level (C.L.).

Including the much larger population of distant repeating FRBs, which cannot be excluded on an individual basis, produces a 95\% C.L.\ upper limit on the population density of such FRBs at $z=0$ of $27$\,Gpc$^{-3}$ if repeating FRBs produce emission up to $10^{42}$\,erg, or $540$\,Gpc$^{-3}$ if they consistent of a separate population with a burst energy cut-off of $10^{40}$\,erg. This density is far lower than that of even rare events, such as long duration gamma-ray bursts, or the mergers of white dwarfs, both of which have been proposed as repeating FRB progenitors.

These limits are weakened when considering beamed emission, bursts unobservable in the ASKAP frequency range of 1.128--1.464\,GHz, or (to a lesser extent) anomalously high DMs from intervening matter.

Regardless of either of these considerations, a population of repeating FRBs with properties similar to that of FRB~121102 cannot explain the observed number of single bursts (19) detected in the survey, i.e.\ FRB~121102 is an atypical object. Comparisons with the observed dispersion measure distribution in the lat50 survey favour a much larger population of less-rapidly repeating FRBs, or (less likely) FRBs with a Poissonian distribution of burst arrival times. An alternative explanation is a second population of once-off FRBs.

The ability to use limits to exclude FRB progenitor models is generally hampered by their lack of predictive power. It is hoped that this will change now that the first volumetric limits on the population density of FRBs --- albeit only repeating ones of a certain strength --- have been published. Other experiments surveying for FRBs are also encouraged to use this or a similar approach to derive limits on the repeating FRB population.

\section*{Acknowledgements}

This work was supported by resources provided by The Pawsey Supercomputing Centre with funding from the Australian Government and the Government of Western Australia. Parts of this research  were conducted by the Australian Research Council Centres of Excellence for All Sky Astrophysics (CAASTRO, CE1101020). C.~W.~James acknowledges the help of K.~W.~Bannister, S.~Bhandari, H.~Qiu and R.~M.~Shannon in accessing ASKAP/CRAFT observing times, and R.~M.~Shannon, S.~Oslowski, and J.~X.~Prochaska for comments on the manucript. Calculations in this work use NumPy~1.8.0rc1 \citep{NumPy} and SciPy~v0.13.0b1 \citep{SciPy}.

\bibliographystyle{mnras}
\bibliography{bibtex_entries.bib}

\appendix

\section{Properties of FRB~121102}
\label{sec:repeater}

Properties of the first repeating FRB, FRB~121102, are used to establish a baseline for the emission properties of repeating FRBs, the population density of which is limited in this article. Due to its much more recent discovery, the properties of the second repeating FRB --- FRB 180814.J0422+73 --- are far less well constrained.

Results from the Karl G.\ Jansky Very Large Array (VLA) observations of \citet{2017ApJ...850...76L}, and the Breakthrough Listen observations of \citet{2018ApJ...863....2G} using the Robert C.\ Byrd Green Bank Telescope (GBT), are used to characterise FRB~121102.

\citet{2017ApJ...850...76L} describe a multi-telescope campaign to observed FRB~121102, with nine bursts observed in $80$\,hr of observations. All were detected by the VLA at S-band ($3$\,GHz, 1,024\,MHz bandwidth) during the `late 2016' epoch (34\,hr); none were detected during 6\,hr at C-band (6\,GHz, 2,048\,MHz b/w) during this epoch, nor in the `early 2016' epoch of 10\,hr at L-band (1.4\,GHz, 256\,MHz b/w) and 30\,hr at S-band. The time-averaged rate above the S-band fluence threshold $F_{\rm th}$ of 0.148\,Jy\,ms ($7.4$\,$\sigma$) therefore is $R_0=0.11_{-0.04}^{+0.05}$\,hr$^{-1}$ (error ranges are 68\% C.I.); taking only the S-band observations during the late 2016 period produces a rate $R_0$ of $0.26_{-0.09}^{+0.12}$\,hr$^{-1}$. The cumulative brightness distribution is found to be a power-law:
\begin{eqnarray}
R(F>F_{\rm th}) & = & R_0 \left( \frac{F_{\rm th}}{F_0}\right)^{\gamma} \label{eq:rateobs}
\end{eqnarray}
with $\gamma=-0.7$.

\citet{2018ApJ...863....2G} report a total of 21 bursts in a six hr observation using the 4--8\,GHz receiver of the GBT. The corresponding rate $R$ is $3.5 \pm 0.75$\,hr$^{-1}$; however, 18 of the bursts were observed in the first 0.5\,hr, giving $R=36 \pm 8.5$\,hr$^{-1}$ for this period. While \citet{2018ApJ...863....2G} do not publish a detection threshold, it can be estimated from the signal-to-noise (S/N) values of the shortest-duration observed pulses and the $6\,\sigma$ detection criteria to be approximately $0.015$\,Jy\,ms to bursts at the time resolution of $0.35$\,ms.  \citet{2018ApJ...866..149Z} extend these results by applying a machine learning algorithm to detect $72$ new pulses in the same data set; however, the threshold for each will vary pulse-to-pulse, is therefore ill-defined. The interesting time structure at very short timescales (10-20\,ms) is also not relevant to this work, which is concerned with hour-long timescales.

\subsection{Brightness distribution}

Unlike $R_0$ from equation~(\ref{eq:rateobs}), $\gamma$ can be robustly estimated without consideration of the varying instrumental sensitivity as a function of pulse width and DM. The methods of \citet{2019MNRAS.483.1342J} produce bias-corrected values of $\gamma=-0.61 \pm 0.20$ (c.f.\ $\gamma=-0.7$ from \citet{2017ApJ...850...76L}) and $\gamma=-1.05 \pm 0.23$ respectively.\footnote{The standard deviation in $\gamma$, $\sigma_{\gamma}$, is approximated as $\sigma_{\gamma}=\gamma/\sqrt{N}$.} Since these are mutually consistent, both samples are combined to estimate $\gamma=-0.91 \pm 0.17$, which is rounded to $-0.9 \pm 0.2$ for this work.

The different time and frequency resolutions of \citet{2017ApJ...850...76L} and \citet{2018ApJ...863..150S} change the effective threshold to FRB~121102 pulses, which were observed with a DM of approximately 565\,pc\,cm$^{-3}$ and duration from $0.2$--$2$\,ms. In neither case was the dispersion smearing significant compared to the time resolution, $t_{\rm res}$. Scaling $F_{\rm th}$ by $t_{\rm res}^{0.5}$ gives a VLA threshold from \citet{2017ApJ...850...76L} approximately $30$ times higher than that of GBT observations from \citet{2018ApJ...863..150S}, explaining a rate difference of $20$--$35$ using equation~(\ref{eq:rateobs}) for $\gamma=-0.91 \pm 0.17$. 

Estimating $\gamma$ allows rate comparisons between instruments with different thresholds. The rate range $3.5 \pm 0.75$\,hr$^{-1}$ at the GBT threshold of $0.015$\,Jy\,ms scales to $0.44^{+0.34}_{-0.20}$\,hr$^{-1}$ at the VLA threshold of $0.148$\,Jy\,ms for this range of $\gamma$, which is broadly consistent with the observed range of $0.07$--$0.38$\,hr$^{-1}$.

Here, nominal values of $R_0=0.26_{-0.17}^{+0.12}$\,hr$^{-1}$ at $F_0=0.148$\,Jy\,ms are used, since the sensitivity and times resolution of these observations are much closer to those of the ASKAP/CRAFT lat50 survey described in Section~\ref{sec:craft}.
Further scaling to ASKAP's sensitivity are discussed in Section~\ref{sec:askap_sensitivity}.

\subsection{Intrinsic properties}

The results of \citet{2017ApJ...850...76L} are used to calculate the intrinsic properties of FRB~121102 from the observed properties. Here, bursts from FRB~121102 were resolved in frequency space, so the intrinsic burst energy $E$ can be calculated using:
\begin{eqnarray}
E & = &  \mathcal{E} \frac{4 \pi D_L^2}{1+z}
\label{eq:Fobs}
\end{eqnarray}
for luminosity distance $D_L$, and bandwidth-integrated energy fluence $\mathcal{E}$, analogous to $F_{\rm obs} \Delta \nu_{\rm obs}$ in equation~(\ref{eq:Fobsfar}).

In order to estimate $\mathcal{E}$ (in Jy\,ms\,MHz, i.e.\ $10^{-16}$ erg\,m$^{-2}$, or $10^{-23}$ J\,m$^{-2}$), the bandwidth-average fluence is not used, but rather the Gaussian frequency space fits of \citet{2017ApJ...850...76L}. These profiles are described by the full-width half-maximum (FWHM) and peak amplitude $S_{I,{\rm peak}}$ in the measured $5$\,ms interval (mJy). The integrated energy fluence $\mathcal{E}$ is thus:
\begin{eqnarray}
{\mathcal{E}} & = & \sqrt{2 \pi}  t_{\rm samp} \, \sigma_{\nu} \, S_{I,{\rm peak}} \\
\sigma_{\nu} & = & \frac{\rm FWHM}{2\sqrt{2 \ln2}}
\end{eqnarray} 
which differs very slightly from the estimate provided by \citet{2017ApJ...850...76L}.

The least-significant pulse, 57638.49937435, was observed with FWHM of 420\,MHz and $S_{I,{\rm peak}}=130$\,mJy, giving $\mathcal{E}=24.6$\,Jy\,ms\,MHz. Given its signal-to-noise of $12\,\sigma$ against a threshold of $7.4\,\sigma$, this implies a threshold value $\mathcal{E}_{\rm th}\approx 76 \cdot 10^3$\,Jy\,s\,Hz, or $7.6 \cdot 10^{-15}$\,erg\,m$^{-2}$.

Using $z=0.19273$ \citep{2017Natur.541...58C}, a flat frequency dependence, and the cosmology of equation~(\ref{eq:cosmology}), the luminosity distance $D_L$ of FRB~121102 is 972\,Mpc, with rate is 7.4\,day$^{-1}$ above an isotropic energy of $E_0=1.7 \cdot 10^{38}$\,erg, i.e.\ repeating FRBs are described by:
\begin{eqnarray}
R(E>E_0) & = &  R_0 \left( \frac{E}{E_0} \right)^{\gamma} \label{eq:intrinsic_rate_w_err} \\
R_0 & = & 7.4_{-4.8}^{+4.0}\,{\rm day}^{-1} \nonumber \\
E_0 & = & 1.7 \cdot 10^{38}\,{\rm erg}  \nonumber \\
\gamma & = & -0.9 \pm 0.2.
\end{eqnarray}
The FWHM is also used to characterise the bandwidth of each burst, i.e.\ $\Delta \nu_{\rm FRB}=420$\,MHz.

The burst durations of \citet{2017ApJ...850...76L} were typically 2\,ms, although much shorter intrinsic burst durations are possible due to the time resolution of the observations. \citet{2018ApJ...863....2G} observe with much higher time resolution, finding bursts lasting as little as $\sim0.2$\,ms. Burst durations of $\Delta t_{\rm FRB}=0.2$--2\,ms are therefore assumed.

\bsp	
\label{lastpage}
\end{document}